\documentclass[sigconf, screen, nonacm]{acmart}

\usepackage{enumitem}
\setlist[itemize]{leftmargin=*}
\usepackage{dirtytalk}

\AtBeginDocument{%
  }

\setcopyright{acmlicensed}
\copyrightyear{2018}
\acmYear{2018}
\acmDOI{XXXXXXX.XXXXXXX}

\acmConference[Conference acronym 'XX]{Make sure to enter the correct
  conference title from your rights confirmation emai}{June 03--05,
  2018}{Woodstock, NY}
\acmBooktitle{Companion Proceedings of the 33rd ACM Symposium on the Foundations of Software Engineering (FSE '25), June 23--27, 2025, Trondheim, Norway}
\acmISBN{978-1-4503-XXXX-X/18/06}

\begin{document}

\title{The Promise and Pitfalls of WebAssembly:\\Perspectives from the Industry }

\author{Ningyu He}
\affiliation{%
  \institution{The Hong Kong Polytechnic University}
  \country{Hong Kong SAR}
}
\author{Shangtong Cao}
\affiliation{%
  \institution{Beijing University of Posts and Telecommunications}
  \country{China}
}
\author{Haoyu Wang}
\affiliation{%
  \institution{Huazhong University of Science and Technology}
  \country{China}
}
\author{Yao Guo}
\affiliation{%
  \institution{Peking University}
  \country{China}
}
\author{Xiapu Luo}
\affiliation{%
  \institution{The Hong Kong Polytechnic University}
  \country{Hong Kong SAR}
}


\begin{abstract}
As JavaScript has been criticized for performance and security issues in web applications, WebAssembly (Wasm) was proposed in 2017 and is regarded as the complementation for JavaScript.
Due to its advantages like compact-size, native-like speed, and portability, Wasm binaries are gradually used as the compilation target for industrial projects in other high-level programming languages and are responsible for computation-intensive tasks in browsers, \textit{e.g.,} 3D graphic rendering and video decoding.
Intuitively, characterizing in-the-wild adopted Wasm binaries from different perspectives, like their metadata, relation with source programming language, existence of security threats, and practical purpose, is the prerequisite before delving deeper into the Wasm ecosystem and beneficial to its roadmap selection.
However, currently, there is no work that conducts a large-scale measurement study on in-the-wild adopted Wasm binaries.
To fill this gap, we collect the largest-ever dataset to the best of our knowledge, and characterize the status quo of them from industry perspectives.
According to the different roles of people engaging in the community, \textit{i.e.,} web developers, Wasm maintainers, and researchers, we reorganized our findings to suggestions and best practices for them accordingly.
We believe this work can shed light on the future direction of the web and Wasm.
\end{abstract}

\maketitle

\section{Introduction}
\label{sec:intro}
Since the World Wide Web (WWW) was proposed and implemented in the 90s~\cite{www}, to meet the increasingly diverse needs of users, the hosted contents on online webpages have been transformed from static ones to dynamic ones.
Such feature supports were originally completed by JavaScript, which, however, has been widely criticized by both industry and academia due to its security and efficiency issues~\cite{bring}.
Under this context, WebAssembly (Wasm), as a low-level programming language proposed in 2017, is thriving, which was designed as a complement to JavaScript in web browsers.
According to statistics~\cite{caniuse}, more than 97\% of worldwide web browsers support running binaries written in Wasm in them.
Because Wasm is well suited for achieving computationally intensive tasks, many industrial applications running on the web use Wasm binaries as the computing core, such as Photoshop, AutoCAD, and Google Earth~\cite{media,3D,autocad}.

Despite its huge success, there is still no work that systematically analyzes Wasm binaries that are \textit{actually deployed in the real world}.
Take the most similar and relevant studies as instances. A. Hilbig \textit{et al.}~\cite{empirical} claimed that they conducted a measurement on \textit{real-world Wasm binaries}. However, nearly 80\% of investigated cases are collected from GitHub repos, which are not guaranteed to be used in actual production, and may also contain toy apps written by developers to learn Wasm.
Moreover, M. Musch \textit{et al.}~\cite{wild} have conducted the most similar work.
However, they have only crawled Wasm binaries from the Alexa top 1M, which is a tiny portion, around one-thousandth, of all hosted websites all over the world. Moreover, at the time of their writing, Wasm was still in its infancy stage, which cannot effectively reflect the prosperity that Wasm had reached in recent years.

To fill this gap and conduct a systematic measurement, we focus on the following industrial-related research questions (RQs).

\textit{RQ1: Hosting environment.}
Wasm binaries should be hosted on websites to provide services to visitors. When are these Wasm binaries deployed in the hosting environment? How are these websites distributed around the world? Are these websites malicious and popular? These questions will help readers understand the characteristics of websites that use Wasm binaries to provide services in the real world, uncovering a crucial part of the Wasm ecosystem.

\textit{RQ2: Meta information of Wasm binaries.}
As a low-level binary format, developers compile projects in high-level programming languages, like C, C++, GO, and Rust, to Wasm binaries instead of directly composing them.
What is the source language of these Wasm binaries? What compilation toolchain does the developer use? Is Wasm a size-efficient format?
For developers, having the answers to these questions can better evaluate the technology stack required to develop Wasm binaries. 

\textit{RQ3: Security Threats.}
Many existing works indicate that Wasm binaries can be used by adversaries with malicious intent~\cite{mining}. In addition, some work has pointed out that there may be vulnerabilities in Wasm binaries that can be exploited, leading to process crashes or even privacy leaks in their hosting environment~\cite{risk,everything,wave}.
Are Wasm binaries deployed in real environments malicious? Do they have vulnerabilities that can be exploited?
Knowing the answers to these two questions allows researchers to better evaluate the positive and negative impacts brought by adopting Wasm binaries.

\textit{RQ4: Practical purpose.}
Portability, as one of the important advantages of Wasm, makes Wasm binaries have good versatility, that is, they can be used in various types of applications and can be supported by multiple main-stream browsers.
What are Wasm binaries used for?
If we can answer this question, we can make a targeted judgment on the development direction of Wasm, as well as the efficiency bottleneck that can be used to conduct optimization for browser vendors and tooling developers.

\textbf{This work.}
To answer these RQs, we need to collect as many once-deployed Wasm binaries and their hosting webpages as possible.
To the best of our knowledge, we have collected the largest ever data set, which is composed of more than 4.6K unique real-world deployed Wasm binaries on the Internet.
Taking advantage of heuristic methods, Wasm-specific analyzers, and online services, we depict the status quo of Wasm binaries that are adopted in real-world from various industry perspectives.
\textbf{For promoting community development, we have uploaded the whole dataset in \href{https://drive.google.com/file/d/1HflCNDkIDyqNLFi4uvp6IdZgfJQCz510/view?usp=sharing}{link}}.

The body of the paper is organized as follows. \S\ref{sec:background} introduces some necessary basic knowledge about Wasm. \S\ref{sec:data} details the adopted method to collect the dataset, as well as the overview of the whole dataset. \S\ref{sec:rq1} to \S\ref{sec:rq4} respectively answer the four RQs. 
Then, according to the different roles of people in the Wasm community, \textit{i.e.,} developers, maintainers, and researchers, we give the corresponding suggestions in \S\ref{sec:lesson}.
Finally, the last three sections discuss the limitations, related work, and conclusion of this work, respectively.

\section{Background}
\label{sec:background}
In this section, we will introduce the necessary preliminary knowledge about Wasm to uncover its unique features.

\vspace{0.05in}
\noindent
\textbf{Portability}
As we underlined in \S\ref{sec:intro}, \textit{portability} is one of the greatest strengths of Wasm, which can be explained from two perspectives.
\textit{Wasm can be taken as the compilation target for more than 40 types of high-level programming languages}, including C/C++, Go, and Rust~\cite{multi-language}. Such compilation processes have been supported by mainstream or even official compilation toolchains, \textit{e.g.,} clang~\cite{clang}, TinyGo~\cite{Go}, and rustc~\cite{Rust}.
Moreover, \textit{Wasm binaries are executed in \textit{runtimes}}, which are arch- and OS-agnostic. Wasm runtimes can be integrated into web browsers or installed as standalone software. Hence, Wasm binaries can be spread and executed in web browsers, laptops, or even embedding devices.

\begin{figure}[t]
    \centering
    \includegraphics[width=\columnwidth]{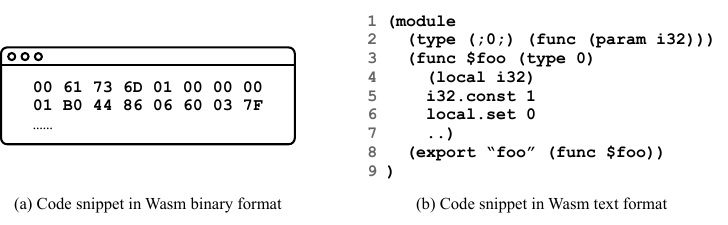}
    \vspace{-0.2in}
    \caption{Difference between binary and text format.}
    \vspace{-0.2in}
    \label{fig:wat-eg}
\end{figure}

\vspace{0.05in}
\noindent
\textbf{Text format \& module.}
As Wasm is an unreadable and compact binary format, it is challenging to perform debugging or code auditing, even for experienced developers. Thus, the official released a toolkit, wabt~\cite{wabt}, which can convert the Wasm binary format and the \textit{Wasm text format} (wat) interchangeably.
Fig.~\ref{fig:wat-eg} illustrates a Wasm binary in both binary and wat format, where we can easily observe the advantage of the wat format in readability.
Specifically, a Wasm binary is often packed in a \textit{module} (L1), consisting of several \textit{sections}, corresponding to different functionalities. For example, the type section at L2 declares a function type that is indexed by \texttt{0}. 
The core of each module is the code section, consisting of function implementations, like \texttt{foo} implemented at L3. Wasm is a \textit{stack-based} language, indicating that operators push operands onto and pop them from the operant stack. For example, \texttt{i32.const 1} intends to push a constant \texttt{1} onto the operant stack.

\vspace{0.05in}
\noindent
\textbf{Data types \& data structures.}
There are four primary data types defined in the Wasm specification, \textit{i.e.,} \texttt{i32}, \texttt{i64}, \texttt{f32}, and \texttt{f64}, where the \texttt{i} and \texttt{f} refer to \textit{integer} and \textit{floating number}, respectively, and the number indicates the length in bits. For instance, \texttt{f32} refers to 32-bit data that is interpreted as a floating number.
Except for the operand stack, other data structures are defined in Wasm, \textit{i.e.,} \textit{local}, \textit{global}, and \textit{memory}.
Specifically, both the local and global follow the mapping structure, mapping from the slot index to the corresponding data, which must be of the four primary types. Note that the local structure can only be accessed within a function, while the global one can be accessed by all functions.
Other non-primary data structures, \textit{e.g.,} string, array, and struct, are stored in the memory area, which can be regarded as a contiguous and byte-addressable array.

\section{Data Collection}
\label{sec:data}
\begin{figure*}[t]
    \centering
    \includegraphics[width=0.7\linewidth]{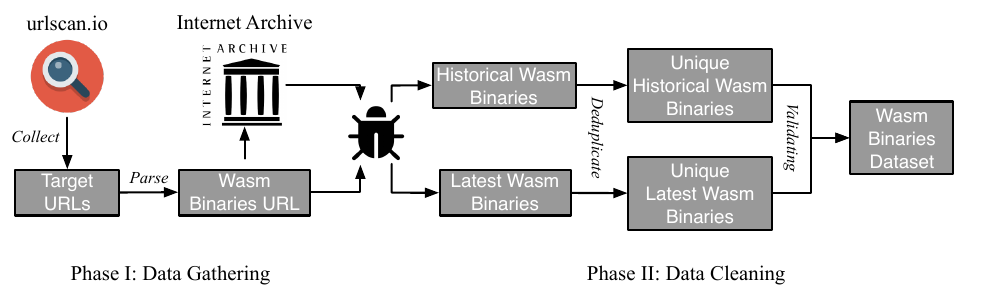}
    \vspace{-0.2in}
    \caption{Workflow of collecting Wasm binaries.}
    \vspace{-0.15in}
    \label{fig:method}
\end{figure*}

To conduct a systematic measurement of real-world deployed Wasm binaries, we intend to collect as many Wasm binaries in the production environment as possible. Thus, we have designed a two-phase collecting process, as shown in Fig.~\ref{fig:method}.

\subsection{Phase I: Data Gathering}
\label{sec:data:phasei}
To collect in-the-wild Wasm binaries, previous work set crawlers targeting the Alexa Top 1M~\cite{wild}. However, the list only covers one-thousandth of all online websites and is not stable enough. For example, nearly 50\% of the websites in the list are replaced every day~\cite{alex}.
Thus, in this work, we take advantage of \textit{urlscan.io}~\cite{urlscan}, a powerful web scanner that can scan over 700K websites and obtain 3.5M files a day and is used by several Fortune 100 companies, governments, and others in the InfoSec industry~\cite{InfoSec}.
Against each website, it behaves as a headless browser, records all traffic data (including all metadata, like transferred file URLs and IPs), and takes a snapshot of the website.
Coupled with its 1B recorded files and 60TB-and-growing storage footprint~\cite{InfoSec}, we can be sure that the scope we covered is dozens or even hundreds of times greater than existing work.

\subsubsection{Obtaining target URLs.}
To ensure that all scanned Wasm binaries are taken into account, we conservatively consider all files ended with \texttt{.wasm} as candidates. Thus, we use the following querying statement in the interface of \textit{urlscan.io}:
\begin{align}\small
\texttt{date: [2017-01-01 TO 2017-01-31] AND filename:*.wasm}\nonumber
\end{align}
to obtain the metadata of all websites that once hosted Wasm-like files monthly.
As Wasm was proposed in 2017, the range of data we collect is from January 2017 to October 2023.
Each website's metadata is packed in the JSON format, where a list records all URLs of transferred files, including JavaScript and Wasm-like files.
Thus, we parse these JSON files to obtain all URLs of transferred files that end with \texttt{.wasm}.

\subsubsection{Crawling latest versions.}
Against these Wasm URLs, we deploy a crawler to retrieve the corresponding files.
We have deployed a proxy pool that includes server nodes in more than a dozen of countries and regions to minimize the false negatives due to network problems. For each file, we set the maximum number of retries to five and randomly switch a proxy if any fetching fails.
It is worth noting that all files we retrieved are the latest versions. In other words, we cannot guarantee the retrieved files are identical to the ones when the urlscan scans.

\subsubsection{Crawling historical versions.}
\label{sec:data:phasei:historical}
Obtaining historical versions can not only restore the actual intent of deploying these files but also observe their longitudinal evolution.
We take advantage of \textit{Internet Archive}, a well-known and widely used web caching service~\cite{archive}, to collect historical cached versions of these Wasm-like files according to the collected Wasm URLs.
Specifically, given a Wasm URL, the Internet Archive may or may not cache it, depending on the caching policy.
If it is, for a URL, a list is kept within the service. Each element of the list is composed of an internal URL for retrieving the cached file, a timestamp, and a digital digest. We only consider the ones that have different digests from the latest one.
We launch the crawler again to obtain these historical versions according to the internal URLs.

\subsection{Phase II: Data Cleaning}
\label{sec:data:phaseii}
Taking advantage of urlscan and Internet Archive, to the best of our knowledge, we have collected the largest-ever dataset of Wasm binaries that are adopted in the real world.
For a precise characterization, some necessary data cleaning is required.

\subsubsection{Removing duplicated historical versions.}
\label{sec:data:phaseii:historical}
Firstly, we need to remove duplicated historical versions collected from Internet Archive.
As we mentioned in \S\ref{sec:data:phasei:historical}, though Internet Archive adopts a digest to distinguish different versions, \textit{different digests do not necessarily mean different file content}.
This is because the digest calculation also considers some metadata~\cite{digest}, which may lead to a digest update even if the file content is unchanged.
Therefore, after crawling files from Internet Archive, we further calculate an MD5 solely based on the file data to eliminate the duplicated ones.

\subsubsection{Removing invalid Wasm binaries.}
\label{sec:data:phaseii:invalid}
We must guarantee that each collected file is a valid Wasm binary, \textit{i.e.,} following the Wasm specification.
Thus, we utilize \textit{wasm2wat}~\cite{wasm2wat}, an official tool that can convert a Wasm binary into the wat format (see \S\ref{sec:background}).
It is enough to verify the validity of Wasm binaries because comprehensive verifications on both syntactic and semantic validity are performed before conducting the conversion~\cite{brewasm}.
Note that we have enabled the \texttt{--enable-all} option during the conversion to avoid false positive alerts due to compatible issues.
Consequently, we regard those files that raise exceptions during the conversion as invalid ones.

\subsection{Data Overview}
\label{sec:data:overview}

\begin{table}[]
\centering
\caption{Basic statistics of collected Wasm binaries.}
\vspace{-0.15in}
\label{table:overview}
\resizebox{0.9\columnwidth}{!}{%
\begin{tabular}{@{}rc@{}}
\toprule
\textbf{\# Webpages\ \ \ }                        & 105,564 \\
\textbf{\# Wasm URL\ \ \ }                        & 112,153 \\
\textbf{\# Unique Wasm URL\ \ \ }                 & 30,424  \\ \midrule
\textbf{\# Latest Wasm Binaries\ \ \ }     & 4,113 (16,436)   \\
\textit{valid}                              & 3,549 (12,789)   \\
\textit{invalid}                            & 564 (3,647)    \\ \midrule
\textbf{\# Historical Wasm Binaries\ \ \ } & 2,170 (3,997)  \\
\textit{valid}                              & 1,057 (2,658)  \\
\textit{invalid}                            & 1,113 (1,339) \\ \bottomrule
\multicolumn{2}{l}{* The numbers in parenthesis are the ones before the deduplicating.}      
\end{tabular}%
}
\vspace{-0.2in}
\end{table}

\subsubsection{URLs Overview}
\label{sec:data:overview:url}
Table~\ref{table:overview} illustrates the basic statistics of collected data.
As we can see, from urlscan, we have identified 105,564 webpages that once hosted Wasm binaries. Among them, we further identified 112,153 Wasm URLs, indicating that some webpages host more than one Wasm binary to achieve complex functionalities. After deduplicating according to URL, 30,424 unique Wasm URLs remain.
When sorting these unique Wasm URLs in descending order of frequency, the top-100 and top-1000 URLs account for more than 43\% and 62\% of all collected Wasm URLs, respectively.
For example, a Wasm URL ending with \texttt{canvaskit.wasm} is included in more than 6.8K webpages. Based on its name and converted wat file, we identify that it is used to help webpages render vector graphics, which is consistent with the unique advantages of Wasm.
This means that the widely referenced Wasm binaries are only a small portion of all existing ones. Their changes, like introducing or deprecating features, could affect the robustness of the services depending on these Wasm binaries.

\subsubsection{Crawled Wasm Binaries}
We have crawled 16,436 latest Wasm binaries in total according to 30K unique Wasm URLs. After deduplicating based on their MD5 values, 4,113 unique ones were finally obtained.
Similarly, we found that the top 100 unique Wasm binaries have been referred to by 10,435 URLs, around 63.5\% (10,435/16,436) of all fetched Wasm binaries.
Interestingly, we found that a file has appeared 2,777 times, where lots of them are fetched according to a set of URLs like \texttt{www.hostingcloud\-.racing/XXX.wasm}. After a manual inspection, this website is marked as a Trojan~\cite{Trojan}.

Moreover, we tried to enlarge the dataset by including once deployed, \textit{i.e.,} historical, Wasm binaries because the crawler can only fetch the latest and current existing ones. Among all 30K unique Wasm URLs, Internet Archive takes snapshots for 843 of them, according to which we have fetched 3,997 historical Wasm binaries in total.
After removing the duplicated historical versions, as we mentioned in \S\ref{sec:data:phaseii:historical}, 2,170 unique historical Wasm binaries remain.
We measured the number of historical versions on a URL basis. We found that the top 100 URLs account for 2,649 snapshots, around 66\% of all snapshots.
Once again, like \S\ref{sec:data:overview:url}, the crawled Wasm binaries prove that the entire Wasm ecosystem is controlled by a small part, which highlights the urgent need to increase the ecological diversity of Wasm.

\begin{table}[]
\centering
\caption{Break-down of invalid reasons.}
\vspace{-0.15in}
\label{table:invalid-reason}
\resizebox{0.8\columnwidth}{!}{%
\begin{tabular}{@{}rccc@{}}
\toprule
\textit{\textbf{}}             & \textbf{Latest} & \textbf{Historical} & \textbf{Total} \\ \midrule
\textit{printable text files}                  & 515             & 101                 & 616            \\
\textit{invalid Wasm binaries} & 4               & 11                  & 15             \\
\textit{other type binaries}   & 30              & 110                 & 140            \\
\textit{unknown binaries}                 & 15              & 891                 & 906            \\ \midrule
\textbf{Total}                 & 564             & 1,113               & 1,677          \\ \bottomrule
\end{tabular}%
}
\vspace{-0.2in}
\end{table}

\subsubsection{Invalid Wasm Binaries}
As we mentioned in \S\ref{sec:data:phaseii:invalid}, we only take the Wasm binaries that do not raise exceptions when invoking wasm2wat as valid ones.
From Table~\ref{table:overview}, we can see that 564 and 1,113 files of the latest and historical categories are marked as invalid, accounting for 13.7\% and 51.2\% of collected ones, respectively.
In other words, we have removed 1,677 invalid cases, and the final dataset is composed of 4,606 unique valid Wasm binaries.

As for the invalid ones, we delve deeper into why exceptions are raised.
We have classified the root causes into four categories, \textit{i.e.,} \textit{printable text files}, \textit{invalid Wasm binaries}, \textit{other type binaries}, and \textit{unknown binaries}.
Specifically, the printable text files category includes all text files, like HTML and JSON files. We believe that the reason why files with the suffix \texttt{.wasm} contain these text files due to \textit{webpage redirect}, where the latter is mistakenly written into the former and obtained by the crawler.
Additionally, the invalid Wasm binaries category indicates the binary has the Wasm magic number, \textit{i.e.,} \texttt{0061736D}~\cite{magic}, but the conversion to the text format fails. Our manual investigation of these cases has observed several interesting situations, \textit{e.g.,} obfuscated code section and self-defined auxiliary sections. We believe that Wasm binaries in this category are more likely to be modified before deployment (like code obfuscation) or compiled from non-standardized compilation toolchains.
As for the cases in the other type binaries, they have valid magic numbers but are of other types of binary, such as DOS MZ executable. We believe that this incorrect extension name annotation is also caused by webpage redirection or deliberate introduction by developers.
At last, we cannot identify any semantics from the cases under the unknown binaries category. After sampling some of them, we believe the most intuitive explanation is that they have been obfuscated comprehensively, including the metadata part.

We can observe that the distribution of invalid reasons of files from different sources is quite different.
For the files directly crawled according to Wasm URLs, most of them are located in the printable text files category. This confirms our speculation that the reason for the appearance of these invalid files is more likely to be due to webpage redirection, which is very common when crawlers obtain webpage content.
As for those collected from the Internet Archive, the vast majority are in the unknown binaries category. As these cases can only be discussed case by case, we have open-sourced them in the dataset to the community for further investigation.

\section{RQ1: Hosting Environment}
\label{sec:rq1}
In this section, we depict the characteristics of those webpages hosted or once hosted \textit{valid} Wasm binaries.
Thus, we consider the webpages with a Wasm URL linked to any of the valid Wasm binaries as candidates in this RQ.
After a filtering process, 78,245 out of 105K webpages are included.

\subsection{Meta Information}
\label{sec:rq1:meta}
We first characterize the time distribution of these webpages monthly.
Although Wasm was proposed in 2017, it was not gradually applied to the production environment until 2021, and experienced explosive growth in 2023. The peak occurred in June 2023, with more than 9,000 new webpages with Wasm binaries appearing this month.
When digging into the reason for the surge, we found a set of URLs appear frequently:
\begin{align}\small
	\texttt{https://meta-bussiness-support-[XXX].firebaseapp.com/}\nonumber
\end{align}
, where they are all unreachable at the time of writing.
We can observe that the word \textit{bussiness} is a typo of \textit{business}; such adoption of typos often appears in phishing scams. We also noticed some forums and security vendors have reported such cases with similar website URLs\footnote{Reddit discussion example: \href{https://www.reddit.com/r/facebook/comments/135f9xa/new_facebook_business_manger_scam/}{link}, security vendor report example: \href{https://www.malwareurl.com/listing.php?domain=meta-business-appealform.firebaseapp.com}{link}.}.
We conclude that this was an organized, large-scale phishing scam controlled by a single group.

\begin{figure}
\centering
\begin{minipage}{.48\columnwidth}
  \centering
  \includegraphics[width=\linewidth]{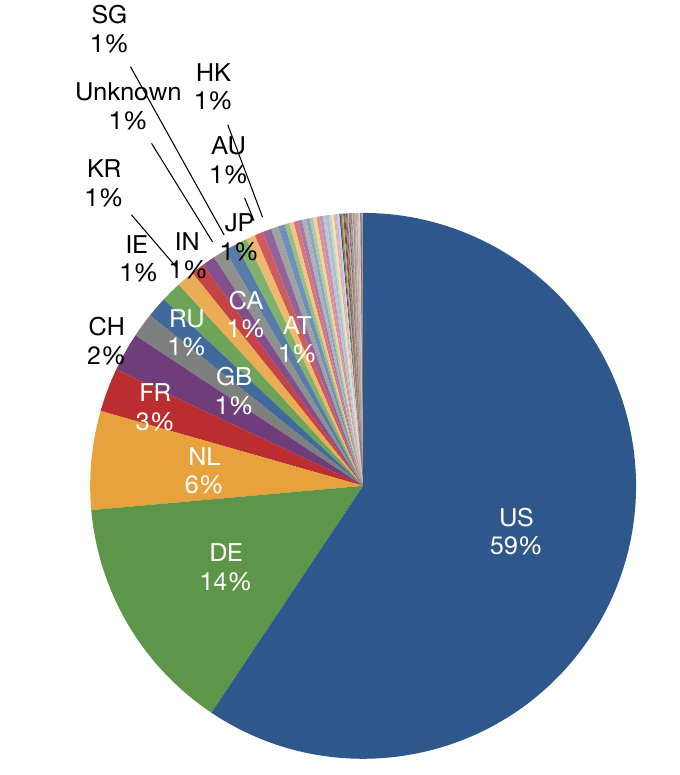}
  \vspace{-0.3in}
  \captionof{figure}{The country distribution of webpages that are hosting or once hosted valid Wasm binaries.}
  \vspace{-0.2in}
  \label{fig:rq1-country}
\end{minipage}%
\hspace{0.2cm}
\begin{minipage}{.48\columnwidth}
  \centering
  \includegraphics[width=\linewidth]{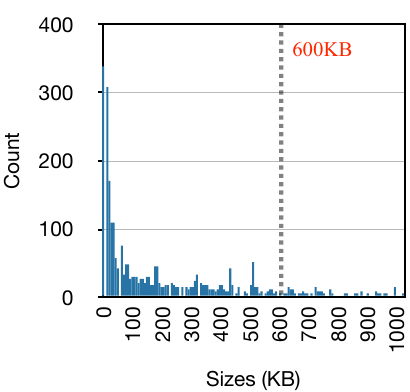}
  \vspace{-0.3in}
  \captionof{figure}{The distribution of sizes of unique real-world Wasm binaries.}
  \vspace{-0.2in}
  \label{fig:rq2-size}
\end{minipage}
\end{figure}

Moreover, according to the IP in the response, we can easily map it to a country or a region to illustrate the geographical distribution of these webpages.
Fig.~\ref{fig:rq1-country} shows the corresponding pie chart.
We can observe that more than half of the websites are deployed in the United States (59\%), followed by five European countries: Germany (14\%), Netherlands (6\%), France (3\%), Switzerland (2\%), and the United Kingdom (1\%).
Countries such as China and India, which have a large number of developers, do not have a large number of Wasm binaries deployed.

\subsection{Popularity}
\label{sec:rq1:popularity}
To measure the popularity of these webpages, we take advantage of the Cisco Umbrella Top 1 Million list~\cite{cisco}\footnote{Cisco Umbrella is a free and objective list that only considers real-world DNS queries instead of other factors, like client IPs and the distribution that are considered by Alexa Top 1 Million~\cite{Alexa}, whose stability is also criticized for replacing nearly 50\% entries within a day~\cite{alex}.}.
Among all 78K webpages, only 13,324 (17.0\%) are included within the top 1M list.
More specifically, only 7 and 171 webpages are included within the top 100 and top 1K, respectively, accounting for 7.0\% and 17.1\%.
In other words, this statistic further illustrates that considering the Wasm binaries only in the top 1M list is not far from enough to describe their overall distribution and ecosystem on the Web.

\subsection{Malice}
\label{sec:rq1:malice}
In \S\ref{sec:rq1:meta}, we have observed some phishing websites. To comprehensively evaluate the malicious intents of these webpages, we decided to take advantage of the \texttt{verdict} field in the scan results provided by urlscan.
Specifically, it is an indicator used to identify the intent of a webpage, where the result is generated jointly by a detection system and user feedback~\cite{verdicts}.
For a given webpage, it will identify it as \textit{legitimate}, \textit{suspicious}, or \textit{malicious}.
For a phishing website, it identifies the brand that the webpage intends to imitate.

To get the most convincing results, \textit{i.e.,} the lower-bound, we only considered webpages that were tagged as \textit{malicious}.
Consequently, out of all 78K webpages, 13,397 are marked as \textit{malicious}, around 17.1\%.
Surprisingly, more than 99\% of them are categorized as \textit{phishing}.
Considering that only millions of phishing webpages are reported out of billions of online ones~\cite{phishing-domain}, we can conclude that \textit{webpages hosting Wasm binaries are more likely to be malicious}.
We then summarized the brands they tried to impersonate, where Facebook, bancolombia, and Interbankpe are the three most commonly impersonated brands, accounting for 69.1\%, 27.9\%, and 2.0\%, respectively, which are confirmed by the taken screenshots.
The large-scale impersonation even partially caused the upward trend of the number of Wasm-hosted webpages in June 2023.
As for the latter two brands, they are both banks that provide financial services located in Colombia and Peru, respectively, which may lead to severe financial losses for their clients.

\vspace{0.1in}
\textit{\textbf{RQ1 Answer:}}
After nearly five years of development, Wasm has only begun to be used in production environments on the Web, though it is far from a consensus to consider Wasm as a front-end language as only 17.0\% webpages once hosted valid Wasm binaries are located within the top-1M list.
Conservatively, up to 17.1\% of Wasm-hosted webpages are related to malicious behavior, especially phishing scams, whose impersonating targets are technical giants and financial service providers.

\section{RQ2: Meta Information of Wasm Binaries}
\label{sec:rq2}
In this section, we illustrate some meta information of 4,606 valid Wasm binaries, including file size, source programming language, and compilation toolchain.
This contributes to the understanding of the whole ecosystem of all adopted Wasm binaries.

\begin{table}[t]
\centering
\caption{String literal keywords to identify source languages}
\vspace{-0.1in}
\label{table:source}
\resizebox{\columnwidth}{!}{%
\begin{tabular}{@{}cl@{}}
\toprule
\textbf{Source language} & \multicolumn{1}{c}{\textbf{String literal keywords}}                              \\ \midrule
C/C++                    & \texttt{C99}, \texttt{C\_plus\_plus}, \texttt{.c}, \texttt{.cpp}, \texttt{clang}, \texttt{clang++}, \texttt{std}, \texttt{emscripten} \\
Rust                     & \texttt{wbg}, \texttt{rustc}, \texttt{.rs}, \texttt{cargo}, \texttt{wasm-bindgen}                          \\
Go                       & \texttt{go}, \texttt{.go}, \texttt{tinygo}                                              \\
C\#                      & \texttt{webcil}                                                        \\ \bottomrule
\end{tabular}%
}
\vspace{-0.2in}
\end{table}

\subsection{Size}
\label{sec:rq2:size}
We first measure the size of all these 4.6K Wasm binaries. The distribution is illustrated in Fig.~\ref{fig:rq2-size}, where we only show the cases whose size is smaller than 1MB due to its long-tail distribution. In total, 2,318 cases are included in this figure, around 50.1\%.
Astonishingly, according to the statistics from HTTP Archive, the median size of requested JavaScript scripts for each site is only around 27.3KB~\cite{httparchive}.
As a binary format, one of Wasm's strengths is its compact size, while we found that no more than 17.7\% of real-world unique Wasm binaries' file size is less than 30KB.
Without considering the uniqueness of these valid Wasm binaries, in the real world, around 66.9\% and 21.4\% of them are smaller than 1MB and 30KB, respectively.
Without considering the uniqueness, we observe an obvious spike at 150KB. It is due to two widely adopted libraries in the Wasm binary format, named \texttt{noise-c.wasm} and \texttt{olm.wasm}, used by 2,555 and 1,026 webpages, respectively.
Both of them are core components in cryptographic algorithms, which is consistent with the unique advantage of Wasm.

\subsection{Source Programming Language \& Compilation Toolchain}
\label{sec:rq2:source}
Identifying the distribution of the source programming language and compilation toolchain of Wasm binaries can deepen the understanding of the whole ecosystem.
As the Wasm binary format cannot reveal the characteristic syntax features in high-level source programming languages, we adopt the following rule-based heuristic strategies.

\textit{Step I: Producers section}. The producers section is a field embedded in Wasm binaries generated by the compiler. Typically, it declares the version of the adopted compiler, which can be used to infer the source programming language.

\textit{Step II: String literal keywords}.
Because not all Wasm binaries have a producers section, we need some heuristics to infer the source language. We thus take advantage of embedded string literals, as all string literals in source files exist in sections of Wasm binaries.
Specifically, the import/export section contains imported/exported library function names, and the data section is composed of string literals used in the program. At last, the custom section may contain debugging information consisting of readable function names and data types.
Therefore, we can extract keywords from these sections to heuristically infer their source language. For example, if the import section declares some APIs of the webcil module, the Wasm binary is compiled from C\#; the data section of a Wasm binary that is compiled from C++ usually begins with \texttt{std::exception} for handling exceptions. All adopted keywords are shown in Table~\ref{table:source}.

\textit{Step III: Voting \& Manual recheck}.
To minimize the possibility of misclassification due to heuristic methods in Step II as much as possible, we design a voting stage.
Specifically, we take import/export, data, and custom section as three voters who independently generate inferring results according to Step II. If two or more voters reach a consensus, \textit{i.e.,} agreeing on the source language of the Wasm binary, we take the language as the result.
Otherwise, we will ask two experts experienced in the field of Wasm to conduct independent identification work based on their domain knowledge and the features revealed in the Wasm binary (like the structure of functions and functionalities of the corresponding decompiled code). If an agreement still cannot be reached, the case will be marked as \textit{unknown} conservatively.

\begin{figure}
\centering
\begin{minipage}{.45\columnwidth}
 \centering
 \includegraphics[width=0.9\linewidth]{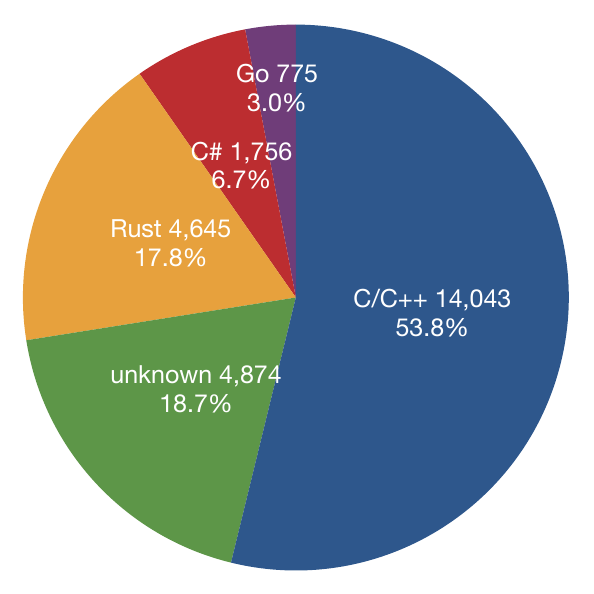}
 \captionof{figure}{The distribution of source languages without considering uniqueness.}
 \vspace{-0.2in}
 \label{fig:rq2-source-no-unique}
\end{minipage}%
\hspace{0.05cm}
\begin{minipage}{.52\columnwidth}
 \centering
 \includegraphics[width=0.9\linewidth]{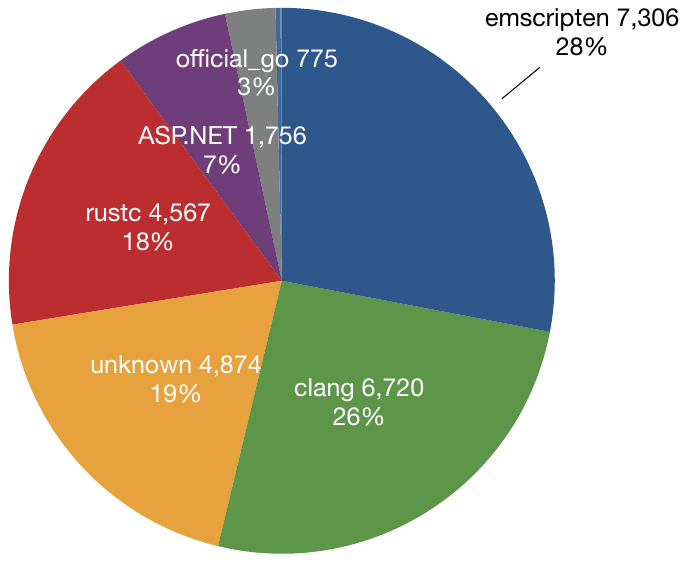}
 \captionof{figure}{The distribution of compilation toolchains without considering uniqueness.}
\vspace{-0.3in}
 \label{fig:rq2-toolchain-no-unique}
\end{minipage}
\end{figure}

\paragraph{Results.}
Consequently, the source language of 303 valid cases (6.6\%) cannot be identified and marked as unknown. The main reason is that some cases are severely obfuscated, so neither textual-level keyword matching nor experts' investigation can effectively determine their source language.
According to the frequency of use, the top three source languages are C/C++, C\#, and Rust, accounting for 46.4\%, 23.7\%, and 15.5\%, respectively.
Without considering the uniqueness, we illustrate the distribution in Fig.~\ref{fig:rq2-source-no-unique}.
We can observe that C/C++ still occupies the first position, accounting for more than 53\% cases. Astonishingly, the number of unknown cases has increased significantly, up to 15 times. We found that the expansion was mainly due to the widespread use of obfuscated Wasm files.

As for the compilation toolchain, we adopt a similar set of strategies. The only difference lies in the second step, where identifying the toolchain requires another set of keywords.
Specifically, different compilation toolchains may import their specialized interfaces that can be used to infer, like \texttt{dlmalloc} adopted by \texttt{emscripten}~\cite{dlmalloc}.
As Fig.~\ref{fig:rq2-toolchain-no-unique} illustrates, we figure out that all Rust and C\# cases are compiled by \texttt{rustc} and \texttt{ASP.NET}, respectively, while the frequency of adopting \texttt{emscripten} and \texttt{clang} is evenly matched. As for the Go cases, less than 3\% are compiled by \texttt{tinygo}, whereas the rest are all compiled by the official Go compiler.
We underline that adopting a different compilation toolchain may negatively impact the portability of Wasm (see \S\ref{sec:background}).
For example, both \texttt{clang} and \texttt{tinygo} support WebAssembly Standard Interface (WASI)~\cite{wasi}, a set of POSIX-like interfaces for operations like file or network accessing, while the official Go compiler does not. Thus, the external environment, \textit{i.e.,} browsers, must either adapt different interfaces for an identical functionality or implement an intermediate layer to support the interfaces generated by different compilation toolchains.

\begin{figure}
\centering
\begin{minipage}{.46\columnwidth}
  \centering
  \includegraphics[width=\linewidth]{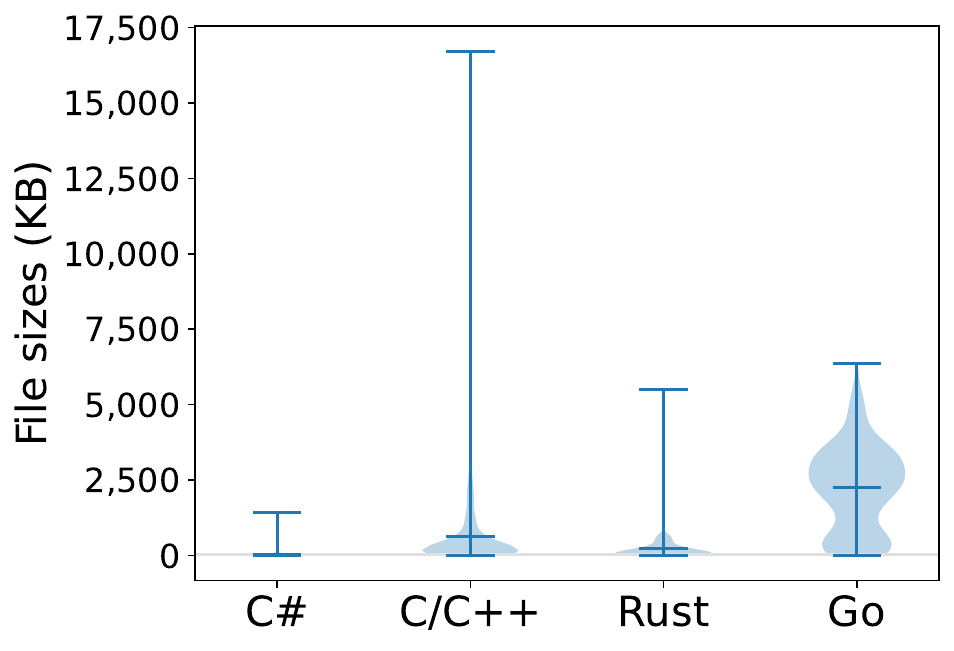}
  \vspace{-0.1in}
  \captionof{figure}{The distribution of file sizes according to source languages.}
  \label{fig:rq2-source-size-relation}
\end{minipage}%
\hfill
\begin{minipage}{.5\columnwidth}
  \centering
  \includegraphics[width=\linewidth]{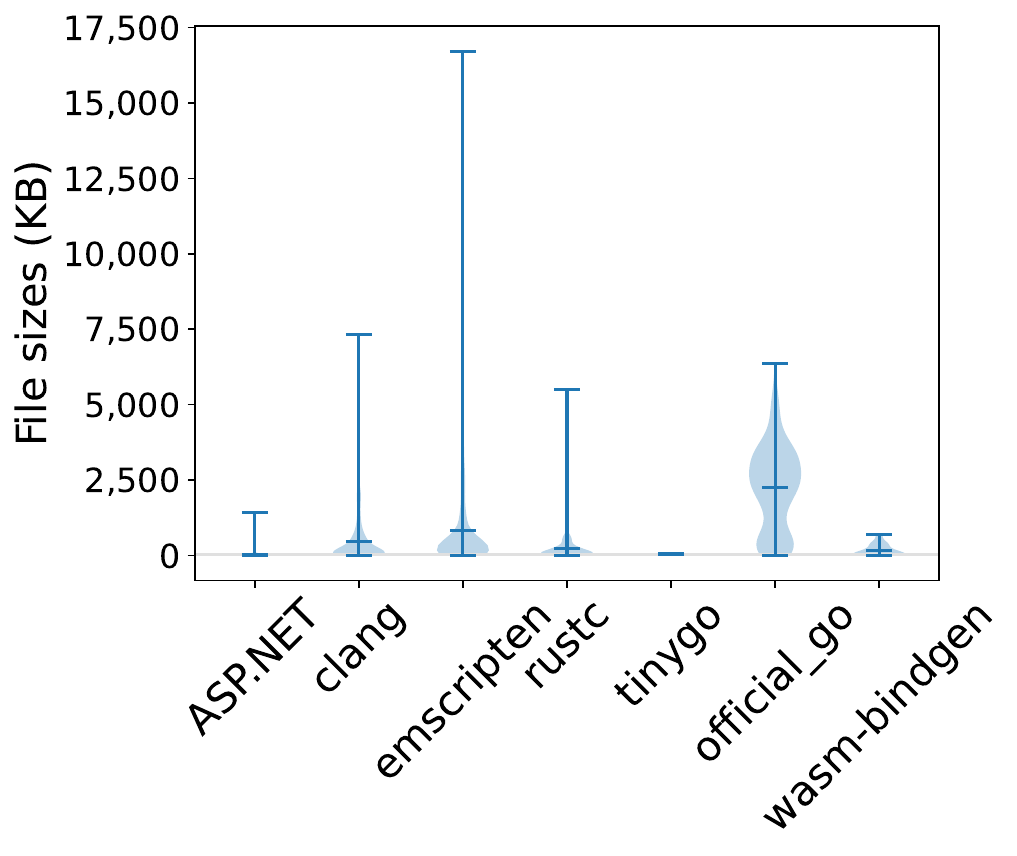}
  \vspace{-0.4in}
  \captionof{figure}{The distribution of file sizes according to compilation toolchains.}
  \label{fig:rq2-toolchain-size-relation}
\end{minipage}%
\hfill
\begin{minipage}{.8\columnwidth}
  \centering
  \includegraphics[width=\linewidth]{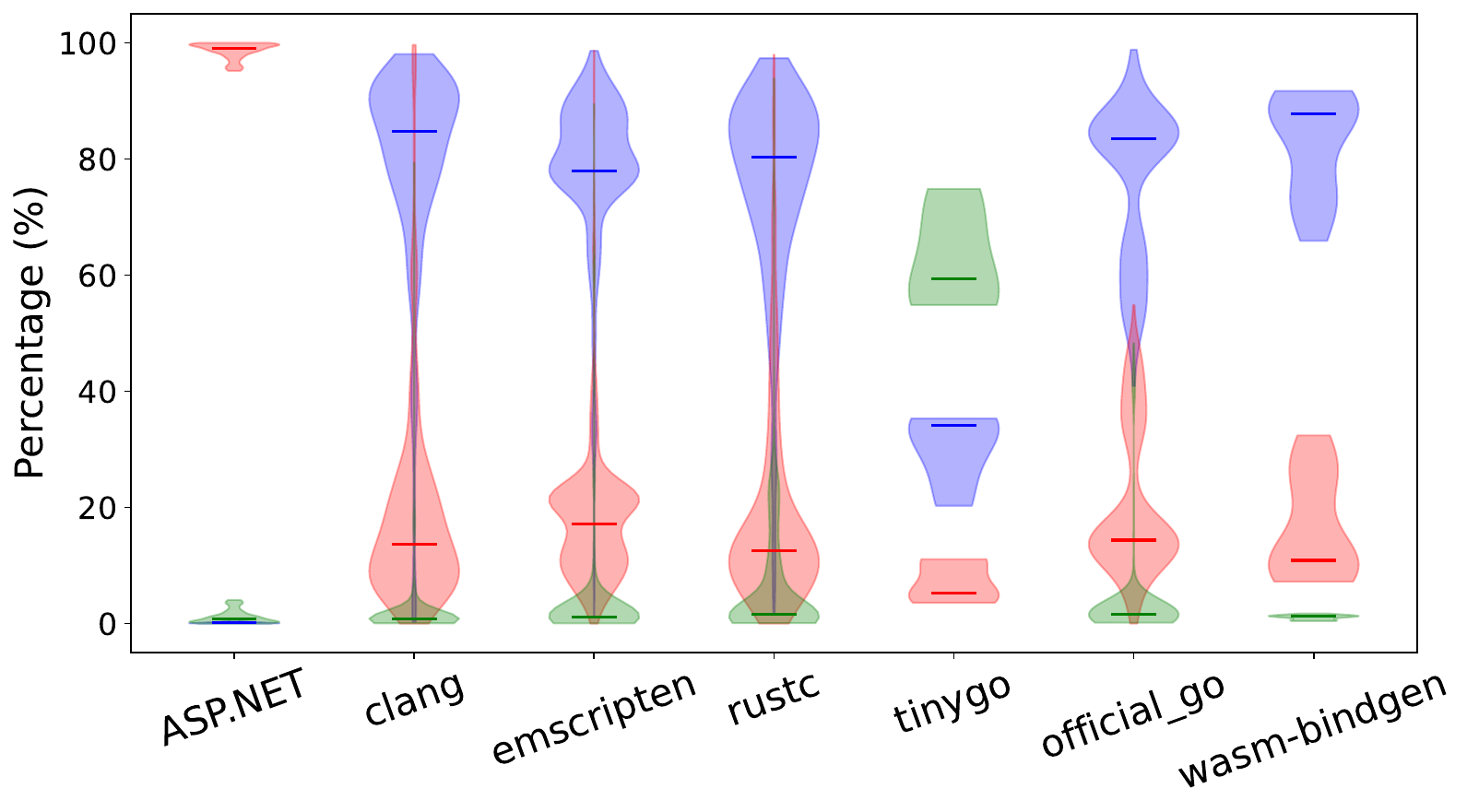}
  \vspace{-0.35in}
  \captionof{figure}{The proportion of sections, where the purple, red, and green refer to the code, data, and other sections, respectively.}
  \vspace{-0.1in}
  \label{fig:rq2-section-ratio}
\end{minipage}
\end{figure}

\subsection{Correlation Analysis}
\label{sec:rq2:cross}
In \S\ref{sec:rq2:size}, we have observed that the majority of real-world adopted Wasm binaries have file sizes larger than average JavaScript scripts', \textit{i.e.,} 27KB. Thus, we conduct a cross-analysis in this section to inspect if there is a correlation between the file size and the source languages/compilation toolchains.

Fig.~\ref{fig:rq2-source-size-relation} illustrates the distribution of file sizes according to the source language, where the horizontal grey line is at 27KB.
As we can see, Wasm binaries compiled from Go have the largest file size on average, and there is a clear peak at 3MB. However, in terms of extreme cases, \textit{i.e.,} outliers with abnormally large file sizes, the situation of C/C++ is the most extreme one. We can see a huge gap between the maximum and average points.
Furthermore, we also depict the distribution of file sizes according to their adopted compilation toolchain, as shown in Fig.~\ref{fig:rq2-toolchain-size-relation}.
As we can see, as for outliers in C/C++, \texttt{emscripten} contributes more than \texttt{clang}. As for Go, all large-size files are generated by the official Go toolchain instead of \texttt{tinygo}.

To explore the root cause of why their file size is so large, we further analyze to which degree each section contributes to the file size. As shown in Fig.~\ref{fig:rq2-section-ratio}, the purple, red, and green parts refer to the code, data, and other sections, respectively.
We can easily observe two strange distributions belonging to \texttt{ASP.NET} and \texttt{tinygo} compilation toolchains.
Specifically, for \texttt{ASP.NET}, we find that the corresponding cases are almost solely composed of the data section. After sampling and manual analysis, we found that developers treat \texttt{ASP.NET} more like a packaging format~\cite{blazor}, which embeds the source code completely into the data section and adopts Wasm binary for transmission.
As for \texttt{tinygo}, we found that sampled cases always have a large custom section consisting of debugging information generated by the compiler. It is worth noting that this section is irrelevant to the semantics when the binary is executed, which can be further deleted to improve the transmission efficiency.
For the remaining Wasm binaries, ranging from KBs to MBs, compiled by other toolchains, around 80\% is related to the code section, \textit{i.e.,} function implementation.
Interestingly, we can observe that the red has penetrated the purple for \texttt{clang}, \texttt{emscripten}, and \texttt{rustc}. In other words, for a small number of cases, \textit{i.e.,} outliers in Fig.~\ref{fig:rq2-source-size-relation}, compiled by these toolchains, the data section instead of the code section contributes the most in terms of file size. This suggests that Wasm binaries are again treated as the data packaging format, like the situation in \texttt{ASP.NET}.
However, this situation is not observed in the official Go compiler, suggesting the huge body of the code section compiled by the official Go toolchain, which lines up with the complaint about the too-many helper functions of Go-compiled Wasm binaries~\cite{go-comp}.

\vspace{0.1in}
\textit{\textbf{RQ2 Answer:}}
Though the compact size is one of the main selling points of Wasm, due to limitations and characteristics brought by programming languages and compilation toolchains, this feature is not fully reflected in the production environment. Only 21.4\% Wasm binaries are smaller than the average size of JavaScript scripts. Moreover, a few languages and toolchains nearly dominate the whole ecosystem, where developers even mistakenly take Wasm as a data packaging format.

\section{RQ3: Security Threats}
\label{sec:rq3}
Security threats, including \textit{malware} and \textit{exploitable vulnerabilities}, in Wasm binaries can introduce significant negative impacts to its hosted webpages, like sensitive data leakage for both developers and visitors.
In this section, we delve deeper into collected Wasm binaries to investigate if they have security threats, \textit{i.e.,} 1) whether they own malicious intents that are introduced by their developers and 2) whether they are vulnerable to certain vulnerabilities that can be exploited by adversaries.

\subsection{Malice}
\label{sec:rq3:malice}
To effectively filter out malware from unreadable Wasm binaries, we take advantage of VirusTotal~\cite{VirusTotal}, a well-known and authoritative scanner specifically designed for malware.
At the time of writing, it consists of 76 vendors. In other words, each case will be scanned by those 76 independent scanners independently, which can minimize false negatives.
Thus, we upload all 4,606 valid Wasm binaries to VirusTotal and download the corresponding report.

Consequently, VirusTotal returns \texttt{413} error code for 179 cases, indicating they are too large for VirusTotal~\cite{VirusTotal}.
For the remaining 4,427 cases, 8 of them are marked as \textit{malicious} and categorized as \textit{cryptominer malware}, which utilizes hardware resources of webpage visitors to mine cryptocurrencies for the webpage owner~\cite{mining}.
After a manual investigation, we think that seven of the reports have high credibility as we can see that the keyword \texttt{cryptonight} exists in multiple sections, like \texttt{\_cryptonight\_hash} in the export section and \texttt{lib/cryptonight/cryptonight.c} in the data section, which is exactly the hashing algorithm used for mining for certain Proof of Work blockchains~\cite{cryptonight}.

Taking advantage of our collected historical versions of Wasm binaries, we can evaluate their evolution.
Finally, we filtered out that 15 Wasm URLs have once hosted these cryptominers. Interestingly, we got a website with multiple different versions archived ranging from Oct. 2017 to Dec. 2017, whose malicious intent can always be detected by VirusTotal from the first version to the latest one.
After a comprehensive search against the URL: \texttt{https://coinhive\-.com/lib/cryptonight.wasm}, we discovered that it is exactly the cryptomining service provider~\cite{coinhive}.
Now, the website is owned by a white hat, who writes a blog titled as \textit{``I Now Own the Coinhive Domain. Here's How I'm Fighting Cryptojacking and Doing Good Things with Content Security Policies.''} pinned on the main page~\cite{how}.
Interestingly, we have observed some developers rename the cryptomining Wasm binaries to confuse the situation, like \texttt{cpuminer\_native.wasm}, \texttt{c.wasm}, and even \texttt{google.wasm}.
Moreover, we have observed that most of these malicious Wasm cryptominers will not be updated after deployment. According to the data we collected, they were all deployed from Dec. 2017 to Jun. 2018, which may be related to the fact that the coinhive's scripts were last updated in 2018.

\subsection{Vulnerability}
\label{sec:rq3:vul}
Wasm can be the compiling target of some high-level programming languages, indicating that vulnerabilities may exist or even be exploitable in Wasm binaries.
A previous study has discussed detecting Wasm binaries vulnerabilities~\cite{issta2023}, which are, however, more like business logic vulnerabilities in specific applications.
Thus, we decided to switch our focus on some general and Wasm-specific vulnerabilities, like the ones focused on by D. Lehmann \textit{et al.}~\cite{everything}.
Specifically, three attack primitives are focused by them, \textit{i.e.,} stack-based buffer overflow, stack overflow, and heap metadata corruption. After summarizing them up, two vulnerabilities are their root causes, that is, \textit{\textbf{unmanaged call stack}} and \textit{\textbf{statically linked allocators}}.
Additionally, other studies have disclosed that importing and invoking some functions can result in exploitations to the hosting environment~\cite{wave}. Thus, we also pay attention to such \textit{\textbf{imported security-related APIs}}.
We characterize these three vulnerabilities in Wasm binaries in this section.

\subsubsection{Unmanaged Call Stack}
\label{sec:rq3:vul:stack}
Unlike managed data, \textit{e.g.,} local and global variables, \textit{unmanaged data} (like call stack and heap) is kept in the linear memory. More importantly, no memory protection mechanisms are adopted, like unmapped pages~\cite{unmap} and ASLR~\cite{aslr}.
Therefore, once attackers obtain write primitives on the \textit{unmanaged call stack}, they can conduct buffer/stack overflow to overwrite data in the heap or even in the data section~\cite{everything}.
In this part, we adopt the heuristic method proposed by D. Lehmann \textit{et al.}~\cite{everything} to filter out: 1) how many binaries maintain such an unmanaged call stack; and 2) how much data is maintained within the stack.
We underline that our analysis is adopted on the largest ever dataset for real-world adopted and non-manually crafted Wasm binaries.

\paragraph{Method.}
Against the first question, we try to determine if a Wasm binary accesses the unmanaged call stack via a stack pointer, which can either be identified by parsing the global section\footnote{If the debug information is kept, the stack pointer will be declared like \texttt{(global \$stack\_pointer (mut i32) (i32.const 66560)}}, or be matched with a heuristic template~\cite{empirical}.
As for the second one, we parse the frame size for each function on the unmanaged call stack by another heuristic template~\cite{everything}.
To complete these two steps, we use APIs exposed by BREWasm~\cite{brewasm}, a Wasm-specific binary rewriter that can translate the binary format into structured objects, to parse each Wasm binary.

\paragraph{Result.}
In total, we have identified 2,777 cases that use the unmanaged call stack, accounting for 60.3\% of all valid cases. Among them, 109 are identified by parsing the global section, while the remaining 2,668 cases are identified by the heuristic method. Astonishingly, without considering the uniqueness and validity, 16,225 out of 20,433 (79.4\%) collected cases use unmanaged call stack.
Then, against 2,777 cases, we enumerate all their functions and try to extract the call frame size. Interestingly, the call frame size can only be extracted from 1,411 cases (50.8\%). In other words, the heuristic cannot match any function of the remaining 1,366 cases.
This partly reflects the possible false positive introduced by the first heuristic, \textit{i.e.,} including binaries that do not use the unmanaged call stack, and the possible false negative imported by the second heuristic because the heuristic may not be general enough for all kinds of source programming languages.
According to the statistics on these 1,411 cases, 35.7\% of functions have accessed the unmanaged call stack, where the average call frame size is 92B, and the median is 16B. We also observed that the call frame of 9.2\% functions is greater than 1KB, indicating a large and free playground for adversaries.

\subsubsection{Statically Linked Allocators}
\label{sec:rq3:vul:allocators}
In Wasm, the linear memory can only be allocated by \texttt{memory.grow} in a coarse-grained and one-directional way, which is inflexible and impractical for complicated functionalities. Thus, compilers may link memory allocators into Wasm binaries to perform fine-grained memory allocations.
Although there are mature memory allocator implementations, considering Wasm's sensitivity to file size, developers may choose to link \textit{lightweight and customized allocators}.
Their implementations may be exploited by malicious users to cause memory-related issues, like the stack-based buffer overflow~\cite{everything}.
Thus, we try to identify all statically linked allocators in collected Wasm binaries.

\paragraph{Method.}
We first extract all readable function names of imported functions from the debugging information. Then, we use the keyword \texttt{alloc} to match related functions. Finally, according to the name, we group them into several specific allocators.
Note that compilers may modify the function names like C++ and Rust always conduct function mangling~\cite{name}. We tried our best to detangle and categorize them into different allocators.

\paragraph{Result.}
We have identified that 338 out of 4,606 valid Wasm binaries have linked a memory allocator in their implementation, 250 of which are \texttt{dlmalloc}, which is the most well-known and -adopted memory allocator. The second place (68) goes to the Go memory allocator~\cite{malloc}, which is specifically designed for Go. We can observe two lightweight allocators, named \texttt{wee\_alloc} (3) and \texttt{emmalloc} (1), respectively. They are specifically designed for the lightweight environment but have proven to have security issues~\cite{empirical}.
For the remaining 16 cases, we cannot determine their specific allocators, though their names and implementations explicitly reveal such an intent. After mapping them back, we can obtain 65 cases with unknown allocators.
We should underline that in a real-world environment, the lack of custom section, \textit{i.e.,} debugging information, will significantly increase the false negative. Thus, we cannot overlook the security risks imported by memory allocators.

\subsubsection{Imported Security-related APIs}
\label{sec:rq3:vul:apis}
Most Wasm binaries are not self-contained, \textit{i.e.,} they rely on some imported functions to achieve their designed functionalities.
Therefore, these imported functions can be taken as attack primitives for adversaries if they are abused. Currently, lots of studies have focused on the security of them~\cite{runtime1,runtime2,runtime3}.
For example, E. Johnson \textit{et al.} have discussed the file system and network isolation from the perspective of the runtime interface~\cite{wave}.
We try to filter out binaries which import such security-related APIs.

\paragraph{Method.}
The basic idea of extracting these imported security-related APIs is through regular-expression-based pattern matching on the readable name of imported functions that are parsed from debugging information.
According to the functionalities of these interface functions, we can divide the security-related ones into the following categories:
\begin{itemize}[leftmargin=*]
\item \textbf{File.} Wasm binaries can access the file system after being authorized. As the file system is actually maintained by the operating system, tampering with it can lead to negative impacts~\cite{file}. Thus, we take the imported functions that include keywords like \texttt{fd} and \texttt{path} into consideration.
\item \textbf{Network.} Wasm binaries can handle network connections, which are standard functionalities provided by WASI~\cite{wasifunc}. For example, \texttt{sock\_recv} can receive data from a socket. These APIs can also lead to some risks, like loading Trojans, if they are abused. Thus, we focus on these \texttt{sock} related APIs.
\item \textbf{Execution.} Except for updating resources maintained by the operating system, Wasm binaries can directly execute commands through interfaces like \texttt{exec} and \texttt{eval}. These commands are notorious and should be paid attention.
\item \textbf{DOM.} As the dataset is collected through web traffic, manipulating the elements in DOM should be paid attention. Abusing these interfaces can lead to traditional web issues, like XSS~\cite{xss}. Thus, we focus on \texttt{document} related imported functions.
\item \textbf{Dynamic linking.} To enrich the functionality and improve the portability of Wasm, Wasm allows dynamic funciton linking at runtime~\cite{dynamic}. This feature is enabled by \texttt{dlopen} and \texttt{dlsym}, which are also included in our pattern library.
\end{itemize}

\begin{table}[]
\centering
\caption{The number of Wasm binaries with imported security-related APIs.}
\vspace{-0.15in}
\label{table:rq3-api}
\resizebox{0.8\columnwidth}{!}{%
\begin{tabular}{@{}r|cc|cc@{}}
\toprule
                         & \textbf{\begin{tabular}[c]{@{}c@{}}\# Unique\\ Wasm\\ Binaries\end{tabular}} & \textbf{\begin{tabular}[c]{@{}c@{}}Ratio\\ (4,606)\end{tabular}} & \textbf{\begin{tabular}[c]{@{}c@{}}\#Wasm\\ Binaries\end{tabular}} & \textbf{\begin{tabular}[c]{@{}c@{}}Ratio\\ (20,433)\end{tabular}} \\ \midrule
\textbf{File}            & 804                                                                        & 17.5\%                                                             & 2,687                     & 13.2\%                                                              \\
\textbf{Network}         & 23                                                                         & 0.5\%                                                              & 55                        & 0.3\%                                                               \\
\textbf{Execution}       & 248                                                                        & 5.4\%                                                              & 1,050                     & 5.1\%                                                               \\
\textbf{DOM}             & 8                                                                          & 0.2\%                                                              & 17                        & 0.1\%                                                               \\
\textbf{\begin{tabular}[c]{@{}c@{}}Dynamic\\ Linking\end{tabular}} & 3                                                                          & 0.1\%                                                              & 9                         & \textless{}0.1\%                                                    \\ \midrule
\textbf{Total}           & \textbf{1,086}                                                             & \textbf{23.6\%}                                                    & \textbf{3,430}            & \textbf{16.8\%}                                                     \\ \bottomrule
\end{tabular}%
}
\vspace{-0.2in}
\end{table}

\paragraph{Results.}
As we can see from Table~\ref{table:rq3-api}, among 4,606 valid and unique Wasm binaries, 1,086 have imported at least one of these security-related APIs, accounting for nearly a quarter. Without considering the uniqueness, the ratio goes to about 16.8\%.
When we break them into categories, we can observe that file- and execution-related APIs are the top 2 factors, accounting for 804 and 248 cases, respectively.
Interestingly, though these Wasm binaries are collected through web traffic, the proportions of the network- and DOM-related ones are negligible.
We speculate this is because these cases are collected from real environments, where developers are likely to remove debugging information or conduct code obfuscation. This would introduce false negatives for the proposed static text-based matching detection.
With current conservative results, we still cannot ignore the possible negative impacts of introducing these APIs on the whole web ecosystem.

\vspace{0.1in}
\textit{\textbf{RQ3 Answer:}}
Security threats in real-world adopted Wasm binaries cannot be neglected. 
On the one hand, Wasm malware is observed, whose malicious intents are covered by changing other seemingly harmless file names.
On the other hand, attack primitives can be obtained by adversaries through accessing unmanaged call stack in functions (35.7\%), exploiting problematic statically linked allocators (65 cases), and invoking imported security-related APIs (23.6\%), which may lead to negative impacts, like XSS attack and sensitive data leakage.
Moreover, we have to underline that these statistics and estimations are \textit{lower-bounds}, as the heavy dependency of syntactic-level features, which may be covered by simply removing debugging information or conducting code obfuscation.

\section{RQ4: Practical Purpose}
\label{sec:rq4}
Wasm has gradually become a programming language used by web developers. Revealing their practical purposes can demystify the web ecosystem from a certain perspective. In this section, we adopt a systematic method to reveal the practical purposes for representative Wasm binaries.

\paragraph{Method.}
Considering the number of datasets we collected, it is unrealistic to analyze the practical purpose of each unique Wasm binary. Therefore, we sorted them in descending order by the number of Wasm binaries represented by each unique sample and selected the Top 100 as candidates.
According to our statistics, the Top 100 account for nearly 15K Wasm binaries.
Then, we converted these 100 cases into the corresponding text format and asked two experts experienced in the Wasm ecosystem to determine their practical purpose. If they disagree on a case, another expert will be introduced to choose one out of the two judgments.
As a control, we also asked ChatGPT-3.5~\cite{gpt} to make the corresponding practical purpose judgments. Specifically, after classifying and summarizing the results of manual analysis, we obtained a total of 6 categories, including ``unknown''. We provided these six categories in Table~\ref{table:rq4}, as well as brief explanations, as options to ChatGPT and asked it to analyze accordingly.
The prompt fed in is as follows:

\say{\textit{You are a full stack engineer. Now, you need to determine the practical purpose of a script. You can start with the import section at the head of the script or the export section and data section at the end. Based on the practical purpose you judge, classify it into one of the following categories: encryption, environment, font processing, graph processing, utilities, and unknown. The brief explanations of these categories are as follows: ... (explanation of each category)}}

\begin{table*}[]
\centering
\caption{Practical purpose of Top-100 the most representative Wasm binaries.}
\vspace{-0.15in}
\label{table:rq4}
\resizebox{.9\textwidth}{!}{%
\begin{tabular}{@{}clcc@{}}
\toprule
\textbf{Category}         & \multicolumn{1}{c}{\textbf{Explanation}}                                                                                                                 & \textbf{Manual} & \textbf{ChatGPT} \\ \midrule
\textbf{encryption}       & \begin{tabular}[c]{@{}l@{}}Libraries or related components used in encryption algorithms, like PRNG and seed generator.\end{tabular}                  & 11              & 15               \\ \midrule
\textbf{environment}      & \begin{tabular}[c]{@{}l@{}}Building a virtual environment to provide emulation for specific hardware or software.\end{tabular}                         & 18              & 6                \\ \midrule
\textbf{\begin{tabular}[c]{@{}c@{}}font processing\end{tabular}}  & \begin{tabular}[c]{@{}l@{}}Dealing with the font-related requests, like rendering fonts.\end{tabular}                                                  & 19              & 20               \\ \midrule
\textbf{\begin{tabular}[c]{@{}c@{}}graph processing\end{tabular}} & \begin{tabular}[c]{@{}l@{}}Dealing with the graph-related requests, like rendering graphs or decoding videos.\end{tabular}                             & 34              & 11               \\ \midrule
\textbf{utilities}        & \begin{tabular}[c]{@{}l@{}}Simple but practical services, like string-matching algorithms and regular expression extractors.\end{tabular} & 13              & 12                \\ \midrule
\textbf{unknown}          & Other situations.                                                                                                                                        & 5               & 36               \\ \bottomrule
\end{tabular}%
}
\vspace{-0.15in}
\end{table*}

\paragraph{Results.}
Table~\ref{table:rq4} illustrates the distribution of the practical purposes of Top-100 Wasm binaries.
As we can see, according to the manual identification results, the \textbf{graph processing} is the dominant one, accounting for more than one-third. This is very consistent with why the Wasm is designed and proposed, \textit{i.e.,} performing computationally intensive work. We have observed cases that are used to decode video streams and render vector graphics.
Interestingly, we observed 19 Wasm binaries are related to \textbf{font processing}. Although they are likely to be different versions of the same service, given the widespread use of each version, we treat them as independent cases. Wasm binaries in this category are mainly responsible for font rendering in the browser, which is still a task that requires graphics processing.
In third place is the \textbf{environment} category, consisting of Wasm binaries taking advantage of its portability to provide a virtual environment for the simulation of different hardware and software. For example, we observed a case dedicated to simulating different game consoles~\cite{emulator}, and some other cases provide runtimes of some programming languages.
The remaining two are \textbf{utilities} and \textbf{encryption} categories. The former provides some simple but effective gadgets, which can also be replaced by JavaScript, while the latter one is composed of computation-intensive encryption-related components.

As for the results generated by ChatGPT, we can observe that more than one-third of cases are marked as \textbf{unknown}.
According to our provided prompt, although we have suggested that ChatGPT can utilize textual information in certain sections, it still cannot make effective judgments in 36 cases.
For the \textbf{encryption} and \textbf{font processing} categories, we compare the results generated by manual efforts and ChatGPT.
The results show that ChatGPT mistakenly categorizes four graphic processing cases into \textbf{encryption} and a string-processing algorithm into \textbf{font processing}.
For the remaining categories, the cases recognized by ChatGPT are always a subset of manual recognition, which reflects the effectiveness and precision of our manual efforts and the limitations of current state-of-the-art AI tools.

\vspace{0.1in}
\textit{\textbf{RQ4 Answer:}}
The practical uses of representative Wasm binaries are mostly consistent with the unique advantages of Wasm, \textit{i.e.,} computational efficiency, and portability. Only 13\% of the Wasm binaries are used for utilities, suggesting its huge growth potential in the web ecosystem.
Moreover, even with the assistance of the most advanced AI currently available, the semantics of Wasm binaries cannot be accurately identified solely by textual information.

\section{Lessons Learned}
\label{sec:lesson}
Based on the systematic data collection and the analysis against four RQs, we further draw suggestions and best practices according to the roles in this community, \textit{i.e.,} web developers, Wasm maintainers, and researchers.

\textbf{For Web Developers.}
Due to the various advantages of Wasm, \textit{e.g.,} efficiency and portability, web developers choose to compile software written in other programming languages into Wasm binaries. For them, we have the following suggestion.
\begin{itemize}
\item \textbf{Choose the appropriate toolchain depending on the situation.} Developers need to adopt a suitable toolchain to take advantage of Wasm's advantages to avoid unnecessary losses in efficiency or security. Specifically, differences in toolchains will mainly affect two aspects: file size and compatibility. As for the former, different compilers or programming languages may lead to completely different file sizes. As for the latter, differences in compilers and statically linked libraries will cause the interface to be non-universal. In summary, we recommend web developers use \texttt{clang}, \texttt{rustc}, and \texttt{tinygo} to compile projects written in the corresponding languages to obtain the smallest binary size and the most compatible Wasm binaries.
\end{itemize} 

\textbf{For Wasm Maintainers.}
Lots of fancy features are continuously introduced by Wasm maintainers. For them, here are some suggestions inspired by our investigations.
\begin{itemize}
    \item \textbf{Put more effort into designing the Wasm Component Model.}
    Wasm Component Model~\cite{component-model} is a broad-reaching architecture for building interoperable Wasm applications, which was still in development. Against the file size inefficiency, the Component Model allows \textit{decoupling} different parts of different functionalities. In other words, some widely-adopted library functions may not need to be packaged in every Wasm binary.
    Moreover, the Component Model is designed to propose an interface that can abstract functions written in different programming languages, allowing communication among them. This is supposed to further improve the portability of Wasm.
    \item \textbf{Consider more attack primitives.} At the beginning of the design, Wasm maintainers' considerations for the attack model were quite simple~\cite{security}. For example, no memory protection mechanism was introduced for such a simple linear memory model to comply with Occam's Razor principle. However, more and more studies show that malicious users can obtain attack primitives through various means, such as problematic memory allocators or imported functions, thus causing memory safety issues. Therefore, we urge maintainers to consider introducing necessary defense mechanisms to the memory model.
\end{itemize}

\textbf{For Researchers.}
Wasm is still an emerging language, which indicates a huge gap that can be put into effort. Therefore, according to the observations and analysis against collected data, here are some possible directions for researchers.
\begin{itemize}
    \item \textbf{Design a tool that can recover the semantics of Wasm binaries.} It is tough to restore the original semantics solely by relying on the text format. Developers can only remove function names and variable types by deleting debugging information but also adopt code or data obfuscators. Our experimental results show that current state-of-the-art AI tools also suffer from high false negatives. Therefore, researchers need to design tools that can restore semantics, \textit{e.g.,} symbolic executors, as prerequisites for subsequent program analysis or AI-related work.
    \item \textbf{Perform security-related research against the Wasm ecosystem.} Although optimization for efficiency is the core of Wasm, the security-related work on Wasm cannot be underestimated. For example, our analysis shows that around 35\% of functions access the unmanaged call stack. If an attacker can obtain attack primitives through problematic memory allocators or imported functions, considering the convenient accessibility of the Internet and portability of Wasm, there will be considerable economic loss. We call on developers to start security-related work on Wasm, such as strengthening memory access or enhancing permission supervision for imported functions.
\end{itemize}

\section{Discussion}
\label{sec:discussion}
Two threats to validity should be discussed.
1) \textbf{Data retrieve scope.}
Although we utilize urlscan, one of the most advanced webpage scanners with the widest coverage, we still cannot guarantee that it can cover all webpages and obtain all Wasm binaries due to the scale of the Internet.
However, to the best of our knowledge, we emphasize that the data set we collected is the largest one so far, which is composed of really deployed Wasm binaries rather than manually compiling projects into Wasm and then conducting the corresponding measurement.
Therefore, we believe this work can provide some meaningful insights for everyone in the Wasm community.
2) \textbf{Technical novelty.}
In this article, we adopted some relatively simple and technically uncomplicated methods to conduct a measurement of the deployed Wasm binaries we collected. Although the technical novelty is a little limited, we still argue that the objects focused on are not covered by previous work, and the insights and suggestions obtained are meaningful.

\section{Related Work}
\label{sec:related}
There exist some measurement studies against Wasm binaries~\cite{empirical, wild, fast, understand,solution}, compilers~\cite{compiler}, and runtimes~\cite{comprehensive,zhang}.
For example, A. Hilbig \textit{et al.} analyzed the security, languages, and use cases on their collected dataset, composed of crawled Wasm binaries and GitHub projects~\cite{empirical}.
Moreover, Y. Yan and A. Jangda conducted a systematic study on the performance of Wasm programs and proposed some potential optimization opportunities for Wasm~\cite{fast, understand}.
When focusing on the security side, M. Kim \textit{et al.} systematically classified various Wasm binary security techniques and methods proposed in the existing literature and discussed future research directions for Wasm binary security~\cite{solution}.
In addition, A. Romano \textit{et al.} analyzed the lifecycle and impact of bugs in Wasm compilers~\cite{compiler}, while Y. Wang and Y. Zhang conducted a systematic analysis of bugs in Wasm runtimes, classifying them and summarizing their repair strategies~\cite{comprehensive,zhang}.

\section{Conclusion}
\label{sec:conclusion}
In this work, we have collected the largest-ever data set composed of in-the-wild adopted Wasm binaries in webpages. Based on that, we have conducted a large-scale measurement from basic statistics to deeper semantic-related characteristics, \textit{i.e.,} abnormal behaviors and practical purposes.
Against four research questions, we have depicted the status quo of the whole ecosystem from different industry perspectives, and convincing and interesting answers were reached.
Last, according to the different roles of people in the community, we reorganized all findings into corresponding suggestions and best practices, which shed light on the development, research, and adoption of Wasm in the future.

\section*{Ethics}
In this work, we have collected Wasm binaries through urlscan and Internet Archive.
All data we collected are exposed to public that can be accessed by any user. Moreover, we strictly followed their terms, \textit{e.g.,} quota of each API everyday and the valid query statements.
We underline that our work does not raise any ethical issues.

% \begin{acks}
% \end{acks}

\bibliographystyle{ACM-Reference-Format}
\bibliography{binaries}


\begin{thebibliography}{61}

%%% the Web address.

\ifx \showCODEN    \undefined \def \showCODEN     #1{\unskip}     \fi
\ifx \showDOI      \undefined \def \showDOI       #1{#1}\fi
\ifx \showISBNx    \undefined \def \showISBNx     #1{\unskip}     \fi
\ifx \showISBNxiii \undefined \def \showISBNxiii  #1{\unskip}     \fi
\ifx \showISSN     \undefined \def \showISSN      #1{\unskip}     \fi
\ifx \showLCCN     \undefined \def \showLCCN      #1{\unskip}     \fi
\ifx \shownote     \undefined \def \shownote      #1{#1}          \fi
\ifx \showarticletitle \undefined \def \showarticletitle #1{#1}   \fi
\ifx \showURL      \undefined \def \showURL       {\relax}        \fi
% invisible to TeX
\providecommand\bibfield[2]{#2}
\providecommand\bibinfo[2]{#2}
\providecommand\natexlab[1]{#1}
\providecommand\showeprint[2][]{arXiv:#2}

\bibitem[(2023)]%
        {autocad}
\bibfield{author}{\bibinfo{person}{}.} \bibinfo{year}{2023}\natexlab{}.
\newblock \bibinfo{title}{AutoCAD Web App}.
\newblock
\newblock
\urldef\tempurl%
\url{https://madewithwebassembly.com/showcase/autocad/}
\showURL{%
\tempurl}


\bibitem[go-(2023)]%
        {go-comp}
 \bibinfo{year}{2023}\natexlab{}.
\newblock \bibinfo{title}{Complaint about large size of {Go-compiled Wasm binaries}}.
\newblock
\newblock
\urldef\tempurl%
\url{https://www.reddit.com/r/golang/comments/15yi9ub/why_is_gos_wasm_so_slow_and_big_compared_with/}
\showURL{%
\tempurl}


\bibitem[Vir(2023)]%
        {VirusTotal}
 \bibinfo{year}{2023}\natexlab{}.
\newblock \bibinfo{title}{{VirusTotal} official website}.
\newblock
\newblock
\urldef\tempurl%
\url{https://www.virustotal.com}
\showURL{%
\tempurl}


\bibitem[www(2024)]%
        {www}
 \bibinfo{year}{2024}\natexlab{}.
\newblock \bibinfo{title}{Definition of {WWW}}.
\newblock
\newblock
\urldef\tempurl%
\url{https://en.wikipedia.org/wiki/World_Wide_Web}
\showURL{%
\tempurl}


\bibitem[{any.run}(2023)]%
        {Trojan}
\bibfield{author}{\bibinfo{person}{{any.run}}.} \bibinfo{year}{2023}\natexlab{}.
\newblock \bibinfo{title}{The analysis of the website www.hostingcloud.racing}.
\newblock
\newblock
\urldef\tempurl%
\url{https://any.run/report/44e8158b003fcac5151ece925d2d1d1160583ef6e21a8100c5c43ae41dfbad46/955837c9-1eb5-42d3-8425-06f770730773}
\showURL{%
\tempurl}


\bibitem[{backblaze}(2023)]%
        {InfoSec}
\bibfield{author}{\bibinfo{person}{{backblaze}}.} \bibinfo{year}{2023}\natexlab{}.
\newblock \bibinfo{title}{Threat Analysis Firm Taps Backblaze in the Fight Against Cybercrime}.
\newblock
\newblock
\urldef\tempurl%
\url{https://www.backblaze.com/cloud-storage/case-studies/urlscan-io}
\showURL{%
\tempurl}


\bibitem[Bian et~al\mbox{.}(2019)]%
        {mining}
\bibfield{author}{\bibinfo{person}{Weikang Bian}, \bibinfo{person}{Wei Meng}, {and} \bibinfo{person}{Yi Wang}.} \bibinfo{year}{2019}\natexlab{}.
\newblock \showarticletitle{Poster: Detecting webassembly-based cryptocurrency mining}. In \bibinfo{booktitle}{\emph{Proceedings of the 2019 ACM SIGSAC Conference on Computer and Communications Security}}. \bibinfo{pages}{2685--2687}.
\newblock


\bibitem[{Bytecode Alliance}(2024)]%
        {component-model}
\bibfield{author}{\bibinfo{person}{{Bytecode Alliance}}.} \bibinfo{year}{2024}\natexlab{}.
\newblock \bibinfo{title}{Documentation of component model}.
\newblock
\newblock
\urldef\tempurl%
\url{https://component-model.bytecodealliance.org}
\showURL{%
\tempurl}


\bibitem[{caniuse}(2024)]%
        {caniuse}
\bibfield{author}{\bibinfo{person}{{caniuse}}.} \bibinfo{year}{2024}\natexlab{}.
\newblock \bibinfo{title}{{caniuse} webpage}.
\newblock
\newblock
\urldef\tempurl%
\url{https://caniuse.com/wasm}
\showURL{%
\tempurl}


\bibitem[Cao et~al\mbox{.}(2023)]%
        {brewasm}
\bibfield{author}{\bibinfo{person}{Shangtong Cao}, \bibinfo{person}{Ningyu He}, \bibinfo{person}{Yao Guo}, {and} \bibinfo{person}{Haoyu Wang}.} \bibinfo{year}{2023}\natexlab{}.
\newblock \showarticletitle{BREWasm: A General Static Binary Rewriting Framework for WebAssembly}. In \bibinfo{booktitle}{\emph{International Static Analysis Symposium}}. Springer, \bibinfo{pages}{139--163}.
\newblock


\bibitem[{Cisco}(2023)]%
        {cisco}
\bibfield{author}{\bibinfo{person}{{Cisco}}.} \bibinfo{year}{2023}\natexlab{}.
\newblock \bibinfo{title}{Cisco Umbrella Top 1 Million list}.
\newblock
\newblock
\urldef\tempurl%
\url{https://s3-us-west-1.amazonaws.com/umbrella-static/index.html}
\showURL{%
\tempurl}


\bibitem[{coinhive}(2023)]%
        {coinhive}
\bibfield{author}{\bibinfo{person}{{coinhive}}.} \bibinfo{year}{2023}\natexlab{}.
\newblock \bibinfo{title}{Github {coinhive} repository}.
\newblock
\newblock
\urldef\tempurl%
\url{https://github.com/cazala/coin-hive}
\showURL{%
\tempurl}


\bibitem[{Emscripten}(2023)]%
        {dlmalloc}
\bibfield{author}{\bibinfo{person}{{Emscripten}}.} \bibinfo{year}{2023}\natexlab{}.
\newblock \bibinfo{title}{The dlmalloc implementation of {Emscripten}}.
\newblock
\newblock
\urldef\tempurl%
\url{https://github.com/emscripten-core/emscripten/blob/main/system/lib/dlmalloc.c}
\showURL{%
\tempurl}


\bibitem[{EmulatorJS}(2023)]%
        {emulator}
\bibfield{author}{\bibinfo{person}{{EmulatorJS}}.} \bibinfo{year}{2023}\natexlab{}.
\newblock \bibinfo{title}{The {EmulatorJS} official website}.
\newblock
\newblock
\urldef\tempurl%
\url{https://emulatorjs.org/}
\showURL{%
\tempurl}


\bibitem[{Go}(2023)]%
        {malloc}
\bibfield{author}{\bibinfo{person}{{Go}}.} \bibinfo{year}{2023}\natexlab{}.
\newblock \bibinfo{title}{The malloc implementation of {Go}}.
\newblock
\newblock
\urldef\tempurl%
\url{https://go.dev/src/runtime/malloc.go}
\showURL{%
\tempurl}


\bibitem[Haas et~al\mbox{.}(2017)]%
        {bring}
\bibfield{author}{\bibinfo{person}{Andreas Haas}, \bibinfo{person}{Andreas Rossberg}, \bibinfo{person}{Derek~L Schuff}, \bibinfo{person}{Ben~L Titzer}, \bibinfo{person}{Michael Holman}, \bibinfo{person}{Dan Gohman}, \bibinfo{person}{Luke Wagner}, \bibinfo{person}{Alon Zakai}, {and} \bibinfo{person}{JF Bastien}.} \bibinfo{year}{2017}\natexlab{}.
\newblock \showarticletitle{Bringing the web up to speed with WebAssembly}. In \bibinfo{booktitle}{\emph{Proceedings of the 38th ACM SIGPLAN Conference on Programming Language Design and Implementation}}. \bibinfo{pages}{185--200}.
\newblock


\bibitem[He et~al\mbox{.}(2023)]%
        {issta2023}
\bibfield{author}{\bibinfo{person}{Ningyu He}, \bibinfo{person}{Zhehao Zhao}, \bibinfo{person}{Jikai Wang}, \bibinfo{person}{Yubin Hu}, \bibinfo{person}{Shengjian Guo}, \bibinfo{person}{Haoyu Wang}, \bibinfo{person}{Guangtai Liang}, \bibinfo{person}{Ding Li}, \bibinfo{person}{Xiangqun Chen}, {and} \bibinfo{person}{Yao Guo}.} \bibinfo{year}{2023}\natexlab{}.
\newblock \showarticletitle{Eunomia: Enabling User-specified Fine-Grained Search in Symbolically Executing WebAssembly Binaries}.
\newblock \bibinfo{journal}{\emph{arXiv preprint arXiv:2304.07204}} (\bibinfo{year}{2023}).
\newblock


\bibitem[Hilbig et~al\mbox{.}(2021)]%
        {empirical}
\bibfield{author}{\bibinfo{person}{Aaron Hilbig}, \bibinfo{person}{Daniel Lehmann}, {and} \bibinfo{person}{Michael Pradel}.} \bibinfo{year}{2021}\natexlab{}.
\newblock \showarticletitle{An empirical study of real-world webassembly binaries: Security, languages, use cases}. In \bibinfo{booktitle}{\emph{Proceedings of the web conference 2021}}. \bibinfo{pages}{2696--2708}.
\newblock


\bibitem[{httparchive}(2023)]%
        {httparchive}
\bibfield{author}{\bibinfo{person}{{httparchive}}.} \bibinfo{year}{2023}\natexlab{}.
\newblock \bibinfo{title}{The page weight report on httparchive}.
\newblock
\newblock
\urldef\tempurl%
\url{https://httparchive.org/reports/page-weight}
\showURL{%
\tempurl}


\bibitem[{Internet Archive}(2023a)]%
        {digest}
\bibfield{author}{\bibinfo{person}{{Internet Archive}}.} \bibinfo{year}{2023}\natexlab{a}.
\newblock \bibinfo{title}{{CDX} digest not accurately capturing duplicates}.
\newblock
\newblock
\urldef\tempurl%
\url{https://archive.org/post/1009990/cdx-digest-not-accurately-capturing-duplicates}
\showURL{%
\tempurl}


\bibitem[{Internet Archive}(2023b)]%
        {archive}
\bibfield{author}{\bibinfo{person}{{Internet Archive}}.} \bibinfo{year}{2023}\natexlab{b}.
\newblock \bibinfo{title}{{Internet Archive} website}.
\newblock
\newblock
\urldef\tempurl%
\url{https://archive.org/}
\showURL{%
\tempurl}


\bibitem[Jangda et~al\mbox{.}(2019)]%
        {fast}
\bibfield{author}{\bibinfo{person}{Abhinav Jangda}, \bibinfo{person}{Bobby Powers}, \bibinfo{person}{Emery~D Berger}, {and} \bibinfo{person}{Arjun Guha}.} \bibinfo{year}{2019}\natexlab{}.
\newblock \showarticletitle{Not so fast: Analyzing the performance of $\{$WebAssembly$\}$ vs. native code}. In \bibinfo{booktitle}{\emph{2019 USENIX Annual Technical Conference (USENIX ATC 19)}}. \bibinfo{pages}{107--120}.
\newblock


\bibitem[Johnson et~al\mbox{.}(2023)]%
        {wave}
\bibfield{author}{\bibinfo{person}{Evan Johnson}, \bibinfo{person}{Evan Laufer}, \bibinfo{person}{Zijie Zhao}, \bibinfo{person}{Dan Gohman}, \bibinfo{person}{Shravan Narayan}, \bibinfo{person}{Stefan Savage}, \bibinfo{person}{Deian Stefan}, {and} \bibinfo{person}{Fraser Brown}.} \bibinfo{year}{2023}\natexlab{}.
\newblock \showarticletitle{WaVe: a verifiably secure WebAssembly sandboxing runtime}. In \bibinfo{booktitle}{\emph{2023 IEEE Symposium on Security and Privacy (SP)}}. IEEE, \bibinfo{pages}{2940--2955}.
\newblock


\bibitem[Kim et~al\mbox{.}(2022)]%
        {solution}
\bibfield{author}{\bibinfo{person}{Minseo Kim}, \bibinfo{person}{Hyerean Jang}, {and} \bibinfo{person}{Youngjoo Shin}.} \bibinfo{year}{2022}\natexlab{}.
\newblock \showarticletitle{Avengers, assemble! Survey of WebAssembly security solutions}. In \bibinfo{booktitle}{\emph{2022 IEEE 15th International Conference on Cloud Computing (CLOUD)}}. IEEE, \bibinfo{pages}{543--553}.
\newblock


\bibitem[Lehmann et~al\mbox{.}(2020)]%
        {everything}
\bibfield{author}{\bibinfo{person}{Daniel Lehmann}, \bibinfo{person}{Johannes Kinder}, {and} \bibinfo{person}{Michael Pradel}.} \bibinfo{year}{2020}\natexlab{}.
\newblock \showarticletitle{Everything old is new again: Binary security of webassembly}. In \bibinfo{booktitle}{\emph{Proceedings of the 29th USENIX Conference on Security Symposium}}. \bibinfo{pages}{217--234}.
\newblock


\bibitem[Lo et~al\mbox{.}(2023)]%
        {file}
\bibfield{author}{\bibinfo{person}{Edward Lo}, \bibinfo{person}{Ningyu He}, \bibinfo{person}{Yuejie Shi}, \bibinfo{person}{Jiajia Xu}, \bibinfo{person}{Chiachih Wu}, \bibinfo{person}{Ding Li}, {and} \bibinfo{person}{Yao Guo}.} \bibinfo{year}{2023}\natexlab{}.
\newblock \showarticletitle{Fuzzing the Latest NTFS in Linux with Papora: An Empirical Study}.
\newblock \bibinfo{journal}{\emph{arXiv preprint arXiv:2304.07166}} (\bibinfo{year}{2023}).
\newblock


\bibitem[MDN(2023)]%
        {Rust}
\bibfield{author}{\bibinfo{person}{MDN}.} \bibinfo{year}{2023}\natexlab{}.
\newblock \bibinfo{title}{{MDN} web docs website}.
\newblock
\newblock
\urldef\tempurl%
\url{https://developer.mozilla.org/en-US/docs/WebAssembly/Rust_to_wasm}
\showURL{%
\tempurl}


\bibitem[Mears(2023)]%
        {3D}
\bibfield{author}{\bibinfo{person}{Jordon Mears}.} \bibinfo{year}{2023}\natexlab{}.
\newblock \bibinfo{title}{How we're bringing Google Earth to the web}.
\newblock
\newblock
\urldef\tempurl%
\url{https://web.dev/case-studies/earth-webassembly}
\showURL{%
\tempurl}


\bibitem[{medium}(2023)]%
        {Alexa}
\bibfield{author}{\bibinfo{person}{{medium}}.} \bibinfo{year}{2023}\natexlab{}.
\newblock \bibinfo{title}{{Cisco Umbrella Releases Free Top 1 Million Sites List}}.
\newblock
\newblock
\urldef\tempurl%
\url{https://medium.com/cisco-shifted/cisco-umbrella-releases-free-top-1-million-sites-list-8497fba58efe}
\showURL{%
\tempurl}


\bibitem[{Microsoft}(2023)]%
        {blazor}
\bibfield{author}{\bibinfo{person}{{Microsoft}}.} \bibinfo{year}{2023}\natexlab{}.
\newblock \bibinfo{title}{{ASP.NET Core Blazor} hosting models}.
\newblock
\newblock
\urldef\tempurl%
\url{https://learn.microsoft.com/en-us/aspnet/core/blazor/hosting-models?view=aspnetcore-8.0}
\showURL{%
\tempurl}


\bibitem[Musch et~al\mbox{.}(2019)]%
        {wild}
\bibfield{author}{\bibinfo{person}{Marius Musch}, \bibinfo{person}{Christian Wressnegger}, \bibinfo{person}{Martin Johns}, {and} \bibinfo{person}{Konrad Rieck}.} \bibinfo{year}{2019}\natexlab{}.
\newblock \showarticletitle{New Kid on the Web: A Study on the Prevalence of WebAssembly in the Wild}. In \bibinfo{booktitle}{\emph{Detection of Intrusions and Malware, and Vulnerability Assessment: 16th International Conference, DIMVA 2019, Gothenburg, Sweden, June 19--20, 2019, Proceedings 16}}. Springer, \bibinfo{pages}{23--42}.
\newblock


\bibitem[{OpenAI}(2023)]%
        {gpt}
\bibfield{author}{\bibinfo{person}{{OpenAI}}.} \bibinfo{year}{2023}\natexlab{}.
\newblock \bibinfo{title}{The models of {OpenAI API}}.
\newblock
\newblock
\urldef\tempurl%
\url{https://platform.openai.com/docs/models/gpt-3-5}
\showURL{%
\tempurl}


\bibitem[Osmani(2023)]%
        {media}
\bibfield{author}{\bibinfo{person}{Addy Osmani}.} \bibinfo{year}{2023}\natexlab{}.
\newblock \bibinfo{title}{Photoshop is now on the web}.
\newblock
\newblock
\urldef\tempurl%
\url{https://medium.com/@addyosmani/photoshop-is-now-on-the-web-38d70954365a}
\showURL{%
\tempurl}


\bibitem[Pearce et~al\mbox{.}(2013)]%
        {runtime1}
\bibfield{author}{\bibinfo{person}{Michael Pearce}, \bibinfo{person}{Sherali Zeadally}, {and} \bibinfo{person}{Ray Hunt}.} \bibinfo{year}{2013}\natexlab{}.
\newblock \showarticletitle{Virtualization: Issues, security threats, and solutions}.
\newblock \bibinfo{journal}{\emph{ACM Computing Surveys (CSUR)}} \bibinfo{volume}{45}, \bibinfo{number}{2} (\bibinfo{year}{2013}), \bibinfo{pages}{1--39}.
\newblock


\bibitem[Rehman et~al\mbox{.}(2014)]%
        {runtime3}
\bibfield{author}{\bibinfo{person}{Amjad Rehman}, \bibinfo{person}{Sultan Alqahtani}, \bibinfo{person}{Ayman Altameem}, {and} \bibinfo{person}{Tanzila Saba}.} \bibinfo{year}{2014}\natexlab{}.
\newblock \showarticletitle{Virtual machine security challenges: case studies}.
\newblock \bibinfo{journal}{\emph{International Journal of Machine Learning and Cybernetics}}  \bibinfo{volume}{5} (\bibinfo{year}{2014}), \bibinfo{pages}{729--742}.
\newblock


\bibitem[Romano et~al\mbox{.}(2021)]%
        {compiler}
\bibfield{author}{\bibinfo{person}{Alan Romano}, \bibinfo{person}{Xinyue Liu}, \bibinfo{person}{Yonghwi Kwon}, {and} \bibinfo{person}{Weihang Wang}.} \bibinfo{year}{2021}\natexlab{}.
\newblock \showarticletitle{An empirical study of bugs in webassembly compilers}. In \bibinfo{booktitle}{\emph{2021 36th IEEE/ACM International Conference on Automated Software Engineering (ASE)}}. IEEE, \bibinfo{pages}{42--54}.
\newblock


\bibitem[statista(2024)]%
        {phishing-domain}
\bibfield{author}{\bibinfo{person}{statista}.} \bibinfo{year}{2024}\natexlab{}.
\newblock \bibinfo{title}{Number of unique phishing sites detected worldwide}.
\newblock
\newblock
\urldef\tempurl%
\url{https://www.statista.com/statistics/266155/number-of-phishing-domain-names-worldwide/}
\showURL{%
\tempurl}


\bibitem[Sti{\'e}venart et~al\mbox{.}(2022)]%
        {risk}
\bibfield{author}{\bibinfo{person}{Quentin Sti{\'e}venart}, \bibinfo{person}{Coen De~Roover}, {and} \bibinfo{person}{Mohammad Ghafari}.} \bibinfo{year}{2022}\natexlab{}.
\newblock \showarticletitle{Security risks of porting c programs to WebAssembly}. In \bibinfo{booktitle}{\emph{Proceedings of the 37th ACM/SIGAPP Symposium on Applied Computing}}. \bibinfo{pages}{1713--1722}.
\newblock


\bibitem[Tajalizadehkhoob(2019)]%
        {alex}
\bibfield{author}{\bibinfo{person}{Samaneh Tajalizadehkhoob}.} \bibinfo{year}{2019}\natexlab{}.
\newblock \bibinfo{title}{The Tale of Website Popularity Rankings: An Extensive Analysis}.
\newblock
\newblock
\urldef\tempurl%
\url{https://labs.ripe.net/author/samaneh_tajalizadehkhoob_1/the-tale-of-website-popularity-rankings-an-extensive-analysis/}
\showURL{%
\tempurl}


\bibitem[{TinyGo}(2023)]%
        {Go}
\bibfield{author}{\bibinfo{person}{{TinyGo}}.} \bibinfo{year}{2023}\natexlab{}.
\newblock \bibinfo{title}{{TinyGo} official docs webpage}.
\newblock
\newblock
\urldef\tempurl%
\url{https://tinygo.org/docs/guides/webassembly/}
\showURL{%
\tempurl}


\bibitem[{Troy Hunt}(2023)]%
        {how}
\bibfield{author}{\bibinfo{person}{{Troy Hunt}}.} \bibinfo{year}{2023}\natexlab{}.
\newblock \bibinfo{title}{Here's How I'm Fighting Cryptojacking and Doing Good Things with Content Security Policies.}
\newblock
\newblock
\urldef\tempurl%
\url{https://www.troyhunt.com/i-now-own-the-coinhive-domain-heres-how-im-fighting-cryptojacking-and-doing-good-things-with-content-security-policies/}
\showURL{%
\tempurl}


\bibitem[{urlscan}(2023a)]%
        {urlscan}
\bibfield{author}{\bibinfo{person}{{urlscan}}.} \bibinfo{year}{2023}\natexlab{a}.
\newblock \bibinfo{title}{{urlscan} official website}.
\newblock
\newblock
\urldef\tempurl%
\url{https://urlscan.io/}
\showURL{%
\tempurl}


\bibitem[{urlscan}(2023b)]%
        {verdicts}
\bibfield{author}{\bibinfo{person}{{urlscan}}.} \bibinfo{year}{2023}\natexlab{b}.
\newblock \bibinfo{title}{The user verdicts and comments section of urlscan}.
\newblock
\newblock
\urldef\tempurl%
\url{https://urlscan.io/blog/2022/02/10/user-verdicts/}
\showURL{%
\tempurl}


\bibitem[wabt(2023a)]%
        {wabt}
\bibfield{author}{\bibinfo{person}{wabt}.} \bibinfo{year}{2023}\natexlab{a}.
\newblock \bibinfo{title}{wabt tool website}.
\newblock
\newblock
\urldef\tempurl%
\url{https://github.com/WebAssembly/wabt}
\showURL{%
\tempurl}


\bibitem[wabt(2023b)]%
        {wasm2wat}
\bibfield{author}{\bibinfo{person}{wabt}.} \bibinfo{year}{2023}\natexlab{b}.
\newblock \bibinfo{title}{wasm2wat tool website}.
\newblock
\newblock
\urldef\tempurl%
\url{https://webassembly.github.io/wabt/doc/wasm2wat.1.html}
\showURL{%
\tempurl}


\bibitem[Wang et~al\mbox{.}(2023)]%
        {comprehensive}
\bibfield{author}{\bibinfo{person}{Yue Wang}, \bibinfo{person}{Zhide Zhou}, \bibinfo{person}{Zhilei Ren}, \bibinfo{person}{Dong Liu}, {and} \bibinfo{person}{He Jiang}.} \bibinfo{year}{2023}\natexlab{}.
\newblock \showarticletitle{A Comprehensive Study of WebAssembly Runtime Bugs}. In \bibinfo{booktitle}{\emph{2023 IEEE International Conference on Software Analysis, Evolution and Reengineering (SANER)}}. IEEE, \bibinfo{pages}{355--366}.
\newblock


\bibitem[{WebAssembly}(2023a)]%
        {wasifunc}
\bibfield{author}{\bibinfo{person}{{WebAssembly}}.} \bibinfo{year}{2023}\natexlab{a}.
\newblock \bibinfo{title}{The {APIs} of {WASI}}.
\newblock
\newblock
\urldef\tempurl%
\url{https://github.com/WebAssembly/WASI/blob/main/legacy/preview1/docs.md}
\showURL{%
\tempurl}


\bibitem[{WebAssembly}(2023b)]%
        {magic}
\bibfield{author}{\bibinfo{person}{{WebAssembly}}.} \bibinfo{year}{2023}\natexlab{b}.
\newblock \bibinfo{title}{The binary format of {WebAssembly}}.
\newblock
\newblock
\urldef\tempurl%
\url{https://webassembly.github.io/spec/core/binary/modules.html#binary-module}
\showURL{%
\tempurl}


\bibitem[{WebAssembly}(2023c)]%
        {dynamic}
\bibfield{author}{\bibinfo{person}{{WebAssembly}}.} \bibinfo{year}{2023}\natexlab{c}.
\newblock \bibinfo{title}{The dynamic linking of {WebAssembly}}.
\newblock
\newblock
\urldef\tempurl%
\url{https://www.wasm.com.cn/docs/dynamic-linking/}
\showURL{%
\tempurl}


\bibitem[{WebAssembly}(2023d)]%
        {wasi}
\bibfield{author}{\bibinfo{person}{{WebAssembly}}.} \bibinfo{year}{2023}\natexlab{d}.
\newblock \bibinfo{title}{Github {WASI} repository}.
\newblock
\newblock
\urldef\tempurl%
\url{https://github.com/WebAssembly/WASI}
\showURL{%
\tempurl}


\bibitem[{WebAssembly}(2023e)]%
        {clang}
\bibfield{author}{\bibinfo{person}{{WebAssembly}}.} \bibinfo{year}{2023}\natexlab{e}.
\newblock \bibinfo{title}{Github {wasi-sdk} repository}.
\newblock
\newblock
\urldef\tempurl%
\url{https://github.com/WebAssembly/wasi-sdk}
\showURL{%
\tempurl}


\bibitem[{WebAssembly}(2023f)]%
        {security}
\bibfield{author}{\bibinfo{person}{{WebAssembly}}.} \bibinfo{year}{2023}\natexlab{f}.
\newblock \bibinfo{title}{The security design of {WebAssembly}}.
\newblock
\newblock
\urldef\tempurl%
\url{https://github.com/WebAssembly/design/blob/main/Security.md}
\showURL{%
\tempurl}


\bibitem[{WebAssembly}(2023g)]%
        {multi-language}
\bibfield{author}{\bibinfo{person}{{WebAssembly}}.} \bibinfo{year}{2023}\natexlab{g}.
\newblock \bibinfo{title}{{WebAssembly} official website}.
\newblock
\newblock
\urldef\tempurl%
\url{https://webassembly.org/getting-started/developers-guide/}
\showURL{%
\tempurl}


\bibitem[wiki(2023a)]%
        {aslr}
\bibfield{author}{\bibinfo{person}{wiki}.} \bibinfo{year}{2023}\natexlab{a}.
\newblock \bibinfo{title}{Address space layout randomization}.
\newblock
\newblock
\urldef\tempurl%
\url{https://en.wikipedia.org/wiki/Address_space_layout_randomization}
\showURL{%
\tempurl}


\bibitem[wiki(2023b)]%
        {unmap}
\bibfield{author}{\bibinfo{person}{wiki}.} \bibinfo{year}{2023}\natexlab{b}.
\newblock \bibinfo{title}{Buffer overflow protection}.
\newblock
\newblock
\urldef\tempurl%
\url{https://en.wikipedia.org/wiki/Buffer_overflow_protection#Canaries}
\showURL{%
\tempurl}


\bibitem[wiki(2023c)]%
        {xss}
\bibfield{author}{\bibinfo{person}{wiki}.} \bibinfo{year}{2023}\natexlab{c}.
\newblock \bibinfo{title}{Cross-site scripting}.
\newblock
\newblock
\urldef\tempurl%
\url{https://en.wikipedia.org/wiki/Cross-site_scripting}
\showURL{%
\tempurl}


\bibitem[wiki(2023d)]%
        {name}
\bibfield{author}{\bibinfo{person}{wiki}.} \bibinfo{year}{2023}\natexlab{d}.
\newblock \bibinfo{title}{Name mangling}.
\newblock
\newblock
\urldef\tempurl%
\url{https://en.wikipedia.org/wiki/Name_mangling}
\showURL{%
\tempurl}


\bibitem[Wu et~al\mbox{.}(2010)]%
        {runtime2}
\bibfield{author}{\bibinfo{person}{Hanqian Wu}, \bibinfo{person}{Yi Ding}, \bibinfo{person}{Chuck Winer}, {and} \bibinfo{person}{Li Yao}.} \bibinfo{year}{2010}\natexlab{}.
\newblock \showarticletitle{Network security for virtual machine in cloud computing}. In \bibinfo{booktitle}{\emph{5th International conference on computer sciences and convergence information technology}}. IEEE, \bibinfo{pages}{18--21}.
\newblock


\bibitem[{XMRIG}(2023)]%
        {cryptonight}
\bibfield{author}{\bibinfo{person}{{XMRIG}}.} \bibinfo{year}{2023}\natexlab{}.
\newblock \bibinfo{title}{Github {xmrig} repository}.
\newblock
\newblock
\urldef\tempurl%
\url{https://github.com/xmrig/xmrig}
\showURL{%
\tempurl}


\bibitem[Yan et~al\mbox{.}(2021)]%
        {understand}
\bibfield{author}{\bibinfo{person}{Yutian Yan}, \bibinfo{person}{Tengfei Tu}, \bibinfo{person}{Lijian Zhao}, \bibinfo{person}{Yuchen Zhou}, {and} \bibinfo{person}{Weihang Wang}.} \bibinfo{year}{2021}\natexlab{}.
\newblock \showarticletitle{Understanding the performance of webassembly applications}. In \bibinfo{booktitle}{\emph{Proceedings of the 21st ACM Internet Measurement Conference}}. \bibinfo{pages}{533--549}.
\newblock


\bibitem[Zhang et~al\mbox{.}(2023)]%
        {zhang}
\bibfield{author}{\bibinfo{person}{Yixuan Zhang}, \bibinfo{person}{Shangtong Cao}, \bibinfo{person}{Haoyu Wang}, \bibinfo{person}{Zhenpeng Chen}, \bibinfo{person}{Xiapu Luo}, \bibinfo{person}{Dongliang Mu}, \bibinfo{person}{Yun Ma}, \bibinfo{person}{Gang Huang}, {and} \bibinfo{person}{Xuanzhe Liu}.} \bibinfo{year}{2023}\natexlab{}.
\newblock \showarticletitle{Characterizing and Detecting WebAssembly Runtime Bugs}.
\newblock \bibinfo{journal}{\emph{arXiv preprint arXiv:2301.12102}} (\bibinfo{year}{2023}).
\newblock


\end{thebibliography}

\end{document}